\documentclass[aps,prd,twocolumn,eqsecnum,showpacs,preprintnumbers,amsmath]{revtex4}

\usepackage{subfigure}
\usepackage{color,graphicx}
\usepackage{amsfonts,amssymb,theorem}
\newcommand{\ma}[1]{\mbox{$\mathcal{#1}$}}

{\theorembodyfont{\upshape}
}
{\theorembodyfont{\upshape}
}
{\theorembodyfont{\upshape}
}
{\theorembodyfont{\upshape}
}
{\theorembodyfont{\upshape}
}
{\theorembodyfont{\upshape}
}

\newcommand{\dalm}{\kern1pt\vbox{\hrule height 0.9pt\hbox{\vrule width
0.9pt\hskip 2.5pt\vbox{\vskip 5.5pt}\hskip 3pt\vrule width 0.3pt}\hrule height
0.3pt}\kern1pt}

\begin{document}
%\thispagestyle{empty}

%<<<<<<<<<<<<< TITLE >>>>>>>>>>>>>>>%
\title{
Can a primordial black hole or wormhole grow as fast as the universe?
}

%<<<<<<<<<<<<< AUTHOR >>>>>>>>>>>>>>>%
\author{$^{1,2}$B.~J.~Carr}
\email{B.J.Carr-at-qmul.ac.uk}
\author{$^{3}$Tomohiro Harada}
\email{harada-at-rikkyo.ac.jp}
\author{$^{4}$Hideki Maeda}
\email{hideki-at-cecs.cl}

%<<<<<<<<<<<<< ADDRESS >>>>>>>>>>>>>>>%

\address{ 
$^{1}$Astronomy Unit, Queen Mary University of London, Mile End Road, London E1 4NS, UK\\
$^{2}$Research Center for the Early Universe, Graduate School of Science, University of Tokyo, Tokyo 113-0033, Japan\\
$^{3}$Department of Physics, Rikkyo University, Tokyo 171-8501, Japan\\
$^{4}$Centro de Estudios Cient\'{\i}ficos (CECS), Casilla 1469, Valdivia, Chile
}

%<<<<<<<<<<<<< DATE >>>>>>>>>>>>>>>%
\date{\today}

%======================================%
%<<<<<<<<<<<<< ABSTRACT >>>>>>>>>>>>>>>%
%======================================%
\begin{abstract} 
This review addresses the issue of whether there are physically realistic self-similar solutions in which a primordial black hole is attached to an exact or asymptotically Friedmann model for an equation of state of the form $p=(\gamma-1)\rho c^2$.
%with $0 \leq \gamma \leq 2$. 
In the positive pressure case ($1 < \gamma <  2$), there is no such solution when the black hole is attached to an exact Friedmann background via a sonic point. However, it has been claimed that there is a one-parameter family of asymptotically Friedmann black hole solutions providing the ratio of the black hole size to the cosmological horizon size is in a narrow range above some critical value.  There are also ``universal'' black holes in which the black hole has an apparent horizon but no event horizon.  It turns out that both these types of solution are only asymptotically {\it quasi}-Friedmann, because they contain a solid angle deficit at large distances, but they are not necessarily excluded observationally. Such solutions may also exist in the $2/3 \le \gamma < \le 1$ case, although this has not been demonstrated explicitly. 
%could still be physically plausible if the fluctuations generating the black holes were set up in a preceding inflationary period. 
In the stiff case ($\gamma =  2$), there is no self-similar solution in an exact background unless the matter turns into null dust before entering the event horizon, which is a contrived and probably unphysical situation.  However, there may be asymptotically quasi-Friedmann solutions without a sonic point which contain universal black holes. In the negative pressure case ($0 <  \gamma <  2/3$), corresponding to a dark-energy-dominated universe, there is a one-parameter family of black hole solutions which are properly asymptotically Friedmann (in the sense that there is no angle deficit) and such solutions may arise naturally in the inflationary scenario. The ratio of the black hole size to the cosmological horizon size must now be  {\it below} some critical value, so the range is more extended than in the positive pressure case and one needs less fine-tuning. 
If one tries to make a black hole which is larger than this, one finds a self-similar solution which connects two asymptotic regions, one being exactly Friedmann and the other asymptotically quasi-Friedmann. This might be regarded as a cosmological wormhole solution providing one defines a wormhole throat quasi-locally in terms of a non-vanishing minimal area on a spacelike hypersurface.
%, which is $0.70$ for $\gamma=1/3$.
%, as well as asymptotically Friedmann solutions which contain contain wormholes or naked singularities. This has the important physical implication that 
The possibility of self-similar black holes in phantom fluids ($\gamma < 0$), where the black hole shrinks as the big rip singularity is approached, or tachyonic fluids ($\gamma >2$) remains unclear. We also consider the possibility of self-similar black hole solutions in a universe dominated by a scalar field. If the field is massless, the situation resembles the stiff fluid case, so any black hole solution is again contrived, although there may still be universal black hole solutions. The situation is less clear if the scalar field is rolling down a potential and therefore massive, as in the quintessence scenario. Although no explicit asymptotically Friedmann black hole solutions  of this kind are known, they are not excluded and comparison with the $0 <  \gamma < 2/3$ perfect fluid case suggests that they should exist if the black hole is not too large. This implies that a black hole might grow as fast as the cosmological horizon in a quintessence-dominated universe in some circumstances, supporting the proposal that accretion onto primordial black holes may have played a role in the production of the supermassive black holes in galactic nuclei. 
% This appears to be the case today and 
%[We calculate the accretion of a black hole either in the period immediately after inflation or at later stages when the dark energy dominated the matter density. SEPARATE PAPER?]
\end{abstract}

\preprint{RESCEU-8/10}
\preprint{CECS-PH-10/02}

%<<<<<<<<<<<<< PACS NUMBER >>>>>>>>>>>>>>>%
\pacs{
04.70.Bw, 97.60.Lf, 04.40.Nr, 04.25.Dm, 95.35.+d
} 
% CECS-PHY-07/??
\maketitle

\section{Historical Introduction} 
%[{\bf hideki: I have corrected the name ``Zel'dovich''}]
Over the last 40 years there has been much interest in how fast a black hole formed in the early universe, when the density is usually radiation-dominated, would grow. As first pointed out by Zel'dovich and Novikov~\cite{zn1967}, a simple Bondi-type accretion analysis suggests that a primordial black hole (PBH) would not grow much at all if it were much smaller than the cosmological horizon at formation but that  it could grow at the same rate as the universe if its initial size were comparable to it.  (The term ``cosmological horizon'' should here be interpreted as the Hubble horizon if the PBHs form after an inflationary period but the particle horizon otherwise.) One might expect the latter situation to apply, since a PBH must be bigger than the Jeans length at formation~\cite{h1971}, so this suggests that any PBH might grow to the horizon mass at the end of the radiation era, which is around $10^{17}M_{\odot}$. Since there is no evidence for such enormous black holes, for a while it was assumed that no PBHs ever formed. 

However, the validity of the Zel'dovich-Novikov calculation is questionable when the black hole size is comparable to the horizon size because it neglects the expansion of the universe and is not fully relativistic. Indeed, the conclusion that a PBH could grow at the same rate as the universe in the radiation-dominated era was disproved by Carr and Hawking~\cite{ch1974}. They demonstrated this by proving that there is no self-similar solution which contains a black 
hole attached to an exact flat Friedmann background via a sonic point (i.e. in which the black hole forms by purely causal processes). The Zel'dovich-Novikov prediction is therefore definitely misleading in this case. Since the PBH must soon fall well below the horizon size, when their argument should be valid, this suggests that PBHs would not grow much at all.

This gave the subject of PBHs a new lease of life and motivated Hawking to consider the quantum effects associated with black holes (since only PBHs could be small enough for these to be significant). Ultimately, this led to his discovery of black hole radiation~\cite{h1975}, so it is ironic that a consideration of PBH accretion led to the conclusion that they evaporate! 

Carr and Hawking also claimed that there are self-similar solutions which are {\it asymptotically} -- rather than {\it exactly} -- Friedmann at large distances from the black hole. However, these correspond to special acausal initial conditions, in which matter is effectively thrown into the black hole at every distance; they do not contain a sonic point because they are supersonic everywhere. Indeed, such solutions exist in the ``dust'' case, when the cosmological fluid is pressureless~\cite{ch1974}. They even found solutions in which the whole universe is in some sense inside a black hole; these are now termed ``universal'' black holes. 

Subsequently, the Carr-Hawking analysis was extended to perfect fluids with equation of state  $p=(\gamma -1)\rho c^2$ where $p$ is the pressure, $\rho$ is the mass density and 
%$1 < \gamma <2$ so that the pressure is positive;
$\gamma$ is a constant ($4/3$ in the radiation case). This is the most general form for a barotropic equation of state compatible with self-similarity. It was proved that the non-existence of self-similar black holes in an exact Friedmann background applies for all values of $\gamma$ in the range  $1$ to $2$~\cite{c1976,bh1978b}. Indeed, 
%it was shown that 
the only physical self-similar solution which can be attached to an exact external Friedmann solution via a sonic point is Friedmann itself; as the radial coordinate decreases, the other solutions either enter a negative mass regime or encounter another sonic point where the pressure gradient diverges~\cite{bh1978b}. As in the radiation case, there are still acausal black hole solutions but these are again supersonic everywhere; the asymptotically Friedmann solutions which reach a sonic point 
%are still physical but they 
represent density perturbations which grow at the same rate as the universe rather than black holes~\cite{cy1990}.

Later it was realized that none of these positive-pressure self-similar solutions are strictly asymptotically Friedmann after all~\cite{mkm2002}:  there is a solid angle deficit at large distances which might in principle show up in the angular diameter test. It would therefore be more accurate to describe them as asymptotically ``quasi-Friedmann''. Such solutions are not excluded observationally, 
%and they could still be physical in the inflationary scenario.
at least for some parameter range, but they are not physically well-motivated.

The attempt to extend the analysis to stiff fluids ($\gamma =2$) led to some controversy. Lin {\it et al.}~\cite{lcf1976} at first claimed that there {\it is}
a self-similar black hole solution in an exact Friedmann background in this case. However, Bicknell and Henriksen~\cite{bh1978a} showed that this conclusion is invalid because Lin {\it et al.} had misidentified the point corresponding to the black hole event horizon. Bicknell and Henriksen did manage to construct a numerical self-similar solution containing a black hole but it required the stiff fluid to turn into null dust at some point. Although this might seem rather contrived, Reed and Henriksen~\cite{rh1980} later found a solution of this kind by generalizing some work of Hacyan~\cite{hac1979}, involving  a self-similar Vaidya model. However, even this model now seems implausible. The only possibility might be universal black hole solutions which are asymptotically quasi-Friedmann.

It is also interesting to consider the growth of a black hole when the density of the universe is dominated by a scalar field, as expected in many cosmological contexts.  If the scalar field is massless (i.e. if there is no scalar potential), then one might expect the same result to apply as in the stiff fluid analysis, since it is well known that a scalar field is equivalent to a stiff fluid provided the gradient is everywhere timelike~\cite{m1988}. Indeed, the conclusion that there is no self-similar non-universal black hole solution in an exact or asymptotically Friedmann background dominated by a massless scalar field is supported by both
numerical studies~\cite{hc2005c} and analytical calculations~\cite{hmc2006}. 

However, the situation is more complicated if there is a scalar potential (i.e. if the scalar field is massive, the mass being associated with second derivative of the potential) and the discovery that the universe is currently accelerating suggests that this may be the case at the present epoch~\cite{supernova}. 
%The similarity assumption then requires that the potential have an exponential form. 
This has led to a study of black hole accretion in quintessence-dominated universes. Indeed, an argument similar to that advocated by Zel'dovich and Novikov has resurfaced in this context in order to explain the origin of the $10^6$ to $10^9 M_{\odot}$ black holes thought to reside in galactic nuclei~\cite{kr1995}. 
While there are several scenarios for the formation of such supermassive black holes, one possibility is that they originated in the early universe 
and grew self-similarly 
%to their present size 
through accretion of  quintessence before cosmological nucleosynthesis~\cite{bm2002,ch2005},
as well as by purely astrophysical processes later. 

Since this proposal is motivated by a Bondi-type argument which neglects the cosmological expansion, it is just as questionable as the original Zel'dovich-Novikov one. 
%This paper exploits the connection between a massless scalar field and a stiff fluid and considers the flaw in the original Lin et al. analysis in more detail.
This has led to the search for self-similar black hole solutions in quintessence-dominated universes. In this case, the similarity assumption requires that the potential have an exponential form. Our analysis in Ref.~\cite{hmc2006} then shows that there is no self-similar solution with a black hole in an exact or asymptotically Friedmann or asymptotically quasi-Friedmann background if the universe is decelerating and no such solution in an exact Friedmann background if it is accelerating. However, this does not prove non-existence in an asymtotically Friedmann or quasi-Friedmann background for the case in which the background is accelerating.
% and asymptotically Friedmann or quasi-Friedmann. 
Kyo {\it et al.}~\cite{khm2008} have shown that
there is a one-parameter family of self-similar asymptotically Friedmann solutions 
in this case, although it is unclear whether they can contain black holes.

The acceleration of the universe can also be explained if its density is dominated by a perfect fluid with $0 < \gamma < 2/3$, so this has motivated us to look for self-similar black hole solutions in this case~\cite{hmc2007,mhc2007}. Such fluids are very different from positive-pressure ones,
since there are no sound-waves (the sound-speed $\sqrt{p/\rho}$ being imaginary),
so one might expect the conclusion about self-similar black hole solutions to be different. We describe such a fluid as ``dark energy'', although this term is sometimes used more generally and may indeed include quintessence itself. 
% to $2$ {\color{red}\bf when it is negligible.})
%{\color{red}\bf [THE DEFINITION OF QUINTESSENCE IS NOT RELATED TO 
%THE VARIABILITY OF GAMMA. LATER I DESCRIBE 
%THE STATUS ABOUT THE TERMINOLOGY.]} 
One might regard quintessence as a form of dark energy in which the parameter $\gamma$, rather than being constant, may vary as the scalar field rolls down its potential.
%but it is different from a $p=(\gamma -1)\mu$ fluid with fixed $\gamma$. The value of $\gamma$ varies 
However, there is an important physical difference between (constant $\gamma$) dark energy and quintessence because there are sound-waves in the latter case,  the sound-speed being the speed of light at short wavelengths~\cite{hmc2006}. 

Since there are no sound-waves for a dark energy fluid, there can be no black hole solutions in an exact Friedmann background. However, as discussed  in Refs.~\cite{hmc2007,mhc2007}, there {\it are} asymptotically Friedmann solutions containing black holes in this case. Indeed, unlike the positive-pressure case, these are genuinely asymptotically Friedmann rather than asymptotically quasi-Friedmann and the associated inhomogeneities may arise naturally in the inflationary scenario.
%(This conclusion may also apply for $\gamma > 2$ or for $\gamma <0$, assuming these cases are physically plausible.)
% and a positive scalar potential and an accelerating universe~\cite{khm2008}. [OK? {\bf hideki: This is correct, see my note}] 
There is a one-parameter family of such solutions but (in contrast to the implication of the Zel'dovich-Novikov argument) they only exist if the black hole is not too large compared to the particle horizon. These solutions are not analogous to the positive-pressure solutions which Carr and Hawking were originally seeking but they might nevertheless be physical.

If one tries to find an asymptotically Friedmann self-similar solution with a black hole which is larger than the upper limit in the $0 < \gamma < 2/3$ situation, one obtains a cosmological wormhole instead. The transition occurs as the black hole apparent horizon approaches the cosmological apparent horizon, after which both horizons disappear. This is in contrast to the $2/3 < \gamma < 2$ case, where the two apparent horizons never merge~\cite{c1976} and one tends to a separate closed universe as the black hole size increases. (However, the separate universe case is not itself self-similar.) 
%{\color{red}\bf[THE FORMER CASE IS FOR SELF-SIMILAR SOLUTIONS BUT THE LATTER CASE IS NOT FOR SELF-SIMILAR SOLUTIONS. SO THE DIRECT COMPARISON DOES NOT MAKE SENSE.]}
In the wormhole solution the metric tends to an asymptotically Kantowski-Sachs form as one approaches the wormhole throat,  this corresponding to a minimum physical radius, and the solution then connects to another asymptotically Friedmannn or asymptotically quasi-Friedmann universe. 

This paper provides a comprehensive discussion of all these solutions. For the most part, we will avoid mathematical technicalities, so the number of equations is minimized. Although the paper is intended as a general review of previous work in this area, we believe that bringing all the cases together is illuminating and leads to some original insights. 
%together it is also contains some original results useful to bring all the different cases together in order to get a proper overview of the subject. 
In successive sections, we  consider positive-pressure fluids, stiff fluids, scalar and quintessence  fields, dark-energy fluids and finally more exotic possibilities (phantom fluids with $\gamma<0$, negative pressure fluids with $2/3<\gamma<1$ and tachyonic fluids with $\gamma>2$).
%For $2/3<\gamma<2$, the conclusion is that there are physically realistic self-similar solutions in which a black hole can grow as fast as the Universe but only in models which are asymptotically quasi-Friedmann. 
%This is not observationally excluded and might arise naturally in inflation.
%{\color{red}\bf it may not be physically realistic}. {\color{red}\bf [CONTRADICTION!]} 
We conclude that there are certainly self-similar asymptotically Friedmann black hole solutions in the $0<\gamma<2/3$ case and there may also be in the quintessence case, so it is interesting that these situations may be observationally favoured at the present epoch.  
In the final section we 
%estimate the amount of black hole accretion during the dark-energy-dominated phases of the universe and 
draw some general conclusions. Two appendices clarify the dimensions of various quantities used in our analysis and the connection between relevant energy conditions. 

It should be stressed that this paper is not intended to be a review of the more general problem of black holes in a cosmological background, although that problem is of great interest in its own right~\cite{einstein,sultana,nayak,kastor,mou,gibbons}. It is also much more narrowly focussed than the earlier review of self-similar solutions in general relativity by Carr and Coley~\cite{cc1999}.  

\section{Positive-pressure fluid case}

\begin{figure}[htbp]
\begin{center}
%\rotatebox{-90}{
\includegraphics[width=0.4\textwidth]{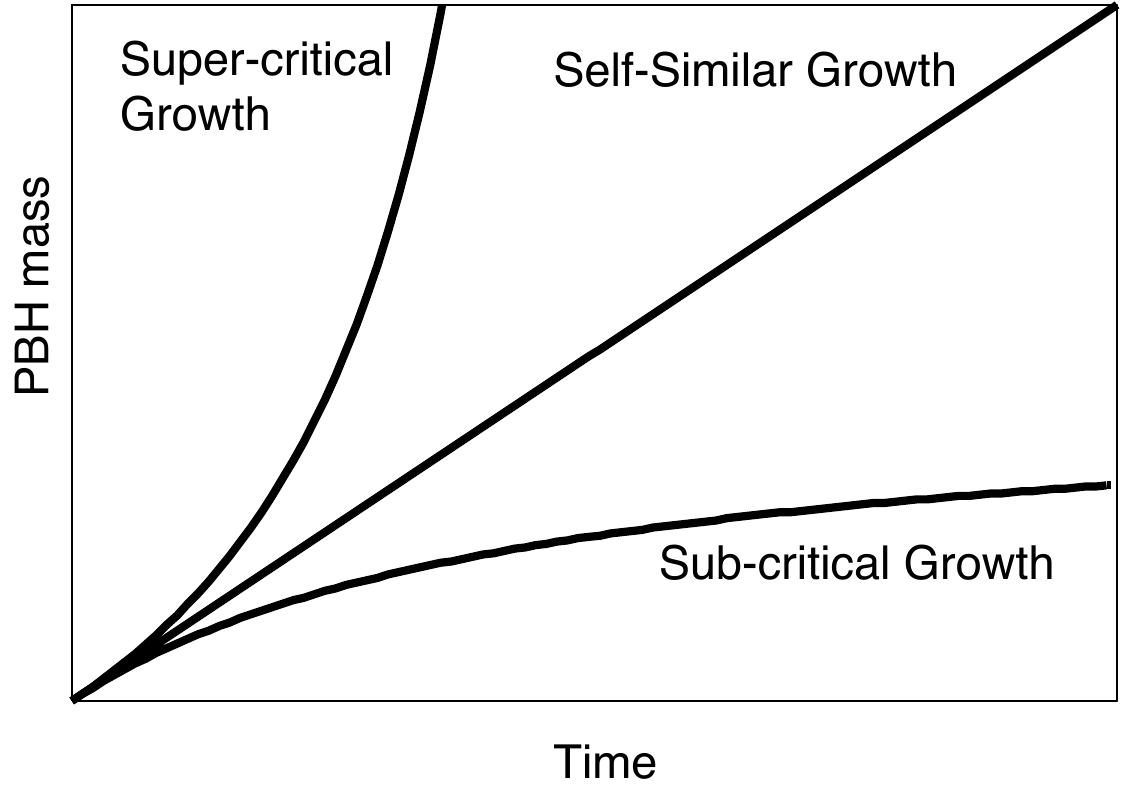}
\includegraphics[width=0.4\textwidth]{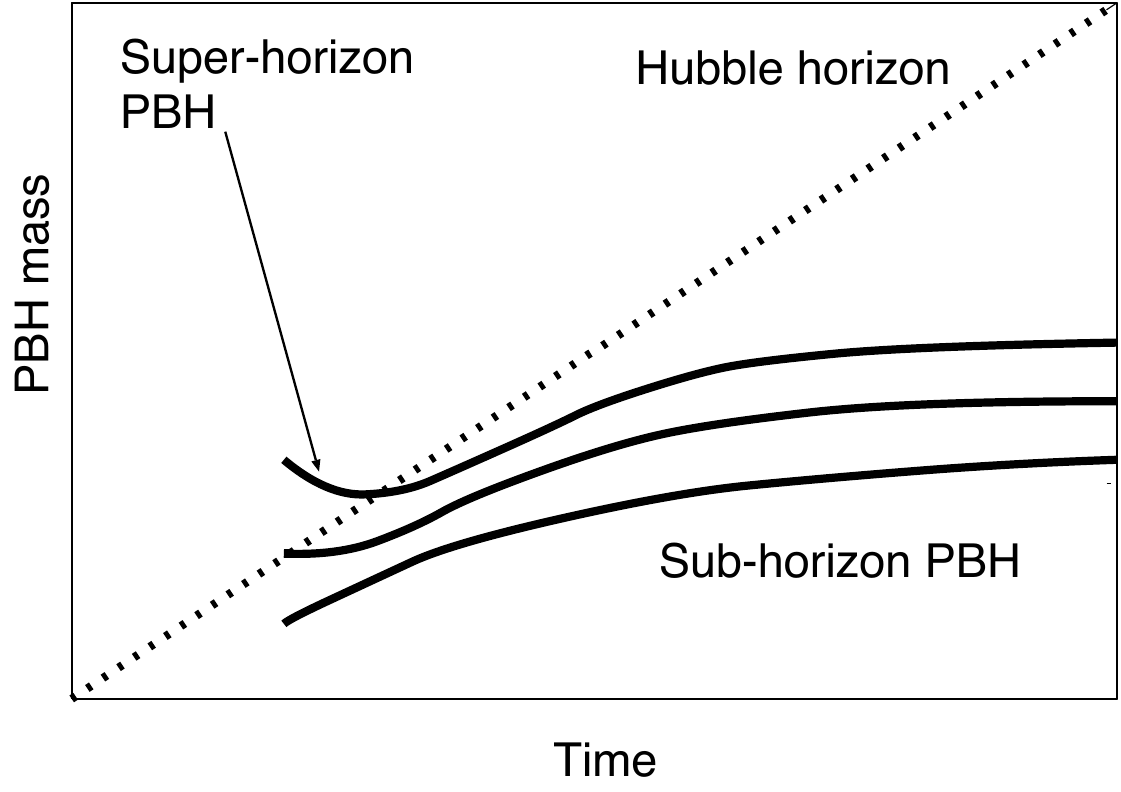}
%}
\caption{\label{accrete}
%$M$ is plotted against $t$ in the $\gamma=1/3$ case for the Newtonian (left) and relativistic (right) analyses.}
Schematic figure showing the growth of PBHs
with positive pressure
for a Bondi-type analysis which neglects the cosmic expansion (top) and a full relativistic analysis which allows for it (bottom).}
\end{center}
\end{figure}

The Zel'dovich-Novikov argument~\cite{zn1967} is based on a Bondi~\cite{b1952} analysis, in which one neglects the cosmological expansion and assumes that the background fluid will be swallowed from within an accretion radius $R_A=2GM/c_s^2$, where $c_s$ is the sound-speed and $M$ is the back hole mass. 
%In the radiation era, $c_s =c/ \sqrt{3}$ , so $R_A$ is just three times the Schwarzschild radius of the black hole. 
The accretion rate for the black hole is then
%of mass $M$ can then be expressed as
\begin{equation}
\frac{dM}{dt} = 4\pi \rho c_s R_A^2 = \frac{16\pi G^2 M^2 \rho}{c_s^3} 
\label{eq:accretion}
\end{equation}
%{\color{red}\bf [THE ABOVE EQUATION IS NOT CORRECT IF WE DISTINGUSH
%THE ACCRETION RADIUS AND THE SCHWARZSCHILD RADIUS.]}
%{\bf [THE ABOVE IS TOO ROUGH IF WE CONSIDER THE EOS DEPENDENCE.]}
where $\rho$ is the mass density at the accretion radius, which is taken to be the background cosmological density. In the radiation-dominated era, which was the case considered by Zel'dovich and Novikov, $c_s =c/ \sqrt{3}$ and so $R_A$ is just three times the Schwarzschild radius of the black hole. Since the mass density in a Friedmann radiation-dominated model is $\rho = 3/(32\pi G t^2)$, where $t$ is the cosmological time from the big bang, Eq.~(\ref{eq:accretion}) then gives
\begin{equation}
\frac{dM}{dt} = \frac{KM^2}{t^2}, \quad K= \frac{9 \sqrt{3}G}{2c^3} \, .
\label{eq:growth}
\end{equation}
Note that this equation cannot be precisely correct.
%because it is based on various approximations. In particular, 
As discussed later, the density near the black hole is not exactly the background value and the accretion radius is not exactly $2GM/c_s^2$, but these effects merely modify the effective value of $K$. Equation~(\ref{eq:growth}) 
%and the expression for $K$ depends on the value adopted for 
can now be integrated to give
\begin{equation}
M=\frac{M_0}{1-\displaystyle{\frac{KM_0}{t_0}\left(1-\frac{t_0}{t}\right)}}\;\; ,
\label{eq:pbhmass}
\end{equation}
where $M_0$ is the mass of the PBH at its formation time $t_0$ and $K^{-1}t_0$ is roughly the mass within the cosmological horizon at that time. (Both the particle horizon and Hubble horizon mass are exactly $c^3t/G$ in the radiation era, which is a factor $9\sqrt{3}/2 \approx 8$ larger.) For $M_0 \ll K^{-1}t_0$,
%, corresponding to an initial black hole mass much less than that of the particle horizon, 
Eq.~(\ref{eq:pbhmass}) implies negligible
accretion. However, for $M_0 = K^{-1}t_0$, it predicts $M= K^{-1}t$, in which case the black hole always has a size comparable to the particle horizon. For $M_0 >  K^{-1}t_0$, the mass  diverges at a time 
\begin{equation}
t_{\infty} = t_0 \left(1- \frac{t_0}{KM_0} \right) ^{-1}
\end{equation}
(i.e. at a time comparable to $t_0$ unless $M_0$ is very close to $K^{-1}t_0$). 
These three behaviours are illustrated by the upper frame in Fig.~\ref{accrete}. 

The middle case corresponds to self-similar growth and is precisely the situation which one might expect to pertain, since an overdense region needs to be larger than the Jeans length at formation but not so large as to form a separate closed universe~\cite{carr1975}, both these length scales being of order the cosmological horizon size. This suggests 
%than
that a PBH might in principle continue to grow as fast as the cosmological horizon until the end of the radiation-dominated era, when its mass would be of order 
$10^{17}M_{\odot}$. The existence of many such holes is excluded, for example, by observations of the large-sale bulk flows of galaxies~\cite{carsak},
%{\color{red}\bf [THE EXISTENCE CANNOT BE EXCLUDED.]}, 
which might lead one to doubt that PBHs ever formed. Of course, the self-similar solution requires very fine-tuning of the black hole mass in the Zel'dovich-Novikov argument but, providing the black holes span a spectrum of masses, one would expect at least some of them to satisfy this condition. 

That the behaviours shown the upper frame in Fig.~\ref{accrete} cannot be correct in a more  precise calculation is indicated by the lower frame in Fig.~\ref{accrete}. This shows the results of fully relativistic numerical calculations for the same initial black hole masses as in the upper frame. We next investigate why the simple Bondi analysis fails and how it can be improved.

\subsection{Improved analysis neglecting cosmic expansion}

The Zel'dovich-Novikov argument can be refined in various ways. In particular, Harada and Carr~\cite{hc2005b} improved the analysis  by allowing the fluid to have a more general barotropic equation of state $p=(\gamma -1)\rho c^2$ with $1<\gamma < 2$ and by including a relativistic focussing factor $\alpha$ in  Eq.~(\ref{eq:accretion}). In this case, the cosmological density is $\rho = 1/(6 \pi G \gamma^2 t^2)$, so Eq.~(\ref{eq:pbhmass}) remains valid but the factor $K$ becomes
\begin{equation}
K=\frac{8G\alpha}{3c^3(\gamma -1)^{3/2}\gamma^2} \, .
\end{equation}
%[WHY DOES THIS GIVE $32/(8\sqrt{3}$ RATHER THAN $3/2$ AT $\gamma = 1/3$?] 
(Their expression was actually a factor of 4 smaller than this because they took the accretion radius to be $GM/c_s^2$ rather than $2GM/c_s^2$.) The length-scale associated with the self-similar black hole solution is
\begin{equation}
R_S =  \frac{2Gt}{c^2 K} = \frac {3 c \gamma^2 ( \gamma -1)^{3/2} t }{4 \alpha} =   \frac {c \gamma ( \gamma -1)^{3/2}}{2 \alpha H} 
\end{equation}
(or a factor $4$ larger in the Harada-Carr analysis), where $H = 2/(3\gamma t)$ is the Hubble parameter.
%Although this does not change the qualitative conclusion of the Zel'dovich-Novikov argument, Harada and Carr 
They also showed that an overdense region is a separate closed universe rather than part of our universe if its size exceeds
\begin{equation}
R_{\rm max} = \frac {2 \sqrt{\pi} \, (3\gamma -2)\, \Gamma \left( \frac{3\gamma -1}{3\gamma -2}\right )c} {3 H  \gamma\, \Gamma \left( \frac{3\gamma}{2(3\gamma -2)}\right)} \, .
\label{maximum}
\end{equation}
This expression  applies for $\gamma > 2/3$ and not just for $1< \gamma <2$. The self-similar scale is always less than the separate-unverse scale in the present analysis but it goes above the self-similar scale for  $\gamma >1.6$ in the original Harada-Carr analysis. They inferred that -- even before attempting a more exact calculation -- one should
have reservations about the self-similar prediction for sufficiently hard equations of state. 

\begin{figure}[htbp]
\begin{center}
%\rotatebox{-90}{
\includegraphics[width=0.5\textwidth]{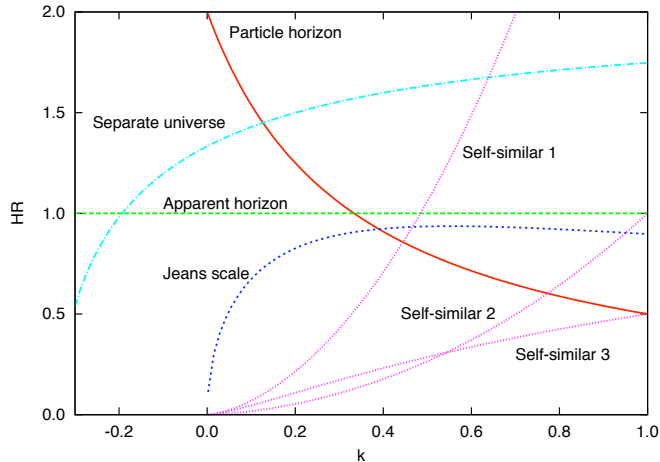}
%}
\caption{\label{scales}
The length scales associated with the Jeans condition, the particle horizon, the Hubble horizon, the separate-universe condition and the self-similar black hole solution are plotted against $k \equiv \gamma-1$. Three self-similar scales are shown, corresponding to the accretion radius being (1) $GM/c_s^2$, (2) $2GM/c_s^2$ and (3) the expression given by Eq.~\eqref{eq:radius}.}
\end{center}
\end{figure}

Harada and Carr gave precise expressions for all relevant  length scales in terms of the equation of state parameter and the functional forms are indicated in Fig.~\ref{scales}. The other scales shown there are the Jeans length~\cite{coles} and the particle horizon size:
\begin{equation}
R_J = \frac {4 \pi c \, \sqrt {\gamma -1}}{(9\gamma -4) H} \, , \quad  R_P = \frac {2c}{(3\gamma -2) H} \, .
\end{equation}
Since there is some uncertainty in the appropriate accretion radius to use, various possibilities for the self-similar scale are shown; these all take the focussing factor $\alpha$ to $1$. Note that some of the scales in Fig.~\ref{scales} can be extended down to $\gamma = 2/3$ and they can all be extended above $\gamma = 2$.

Babichev {\it et al.}~\cite{bde2005} have further refined the accretion analysis by replacing Eq.~(\ref{eq:accretion}) with
 \begin{equation}
\frac{dM}{dt} = \frac{4\pi G^2 AM^2}{c^3} \left( \rho_{\infty} + \frac{p_{\infty}}{c^2} \right),
\label{bab1}
\end{equation}
where the subscript $\infty$ indicates asymptotic values and
 \begin{equation}
 A \equiv  \frac{(3\gamma -2)^{(3\gamma-2)/2(\gamma -1)}}{4 (\gamma-1)^{3/2}}.
 \label{bab2}
  \end{equation}
This allows for the fact that the density of the fluid at the accretion radius is not the asymptotic cosmological value and it also incorporates the relativistic correction to Eq.~(\ref{eq:growth}) associated with the pressure. The Babichev {\it et al.} analysis is not confined to the $p=(\gamma -1)\rho c^2$ equation of state, but we assume this here because of the self-similarity requirement. Thus we again obtain  Eq.~(\ref{eq:growth}) but with $K$ is replaced by
\begin{equation}
K = \frac{2A(\gamma) G}{3\gamma c^3} \, ,
\label{bab3}
\end{equation}
%so one again has an indication of self-similar growth. 
so the mass for self-similar growth now has a more complicated dependence on the equation of state parameter than in the Harada-Carr analysis. As $\gamma$ decreases from 2 to 1, Eq.~(\ref{bab2}) indicates that the factor $A$ increases from 4 to $\infty$, passing through $6\sqrt{3}$ at $\gamma =4/3 $ and diverging like $[e/(\gamma -1)]^{3/2}/4$ as $\gamma \rightarrow 1$. Comparison with Eq.~(\ref{eq:accretion}) also shows that the ``effective'' accretion radius is
 \begin{equation}
 %R_A = \frac{\sqrt{A \gamma} \; GM}{c^{3/2} \sqrt{c_s}} \rightarrow \frac{e^{3/4}GM}{2c_s^2} \approx \frac{GM}{c_s^2} .
 R_A = \frac{GM}{c^2} \sqrt{\frac{A \gamma}{\sqrt{\gamma-1}} } =  \frac{2GM}{c_s^2} \sqrt{ \left( \frac{A}{4} \right) \gamma (\gamma-1)^{3/2}} \, .
 \label{eq:radius}
\end{equation}
The last expression shows the deviation from the naive expression $R_A = 2GM/c_s^2$. For example, this is a factor $\sqrt{2}$ for a stiff fluid and $\sqrt{2/3}$ for a radiation fluid. $R_A$ diverges in the limit $\gamma \rightarrow 1$ and has the form
 \begin{equation}
 R_A \approx \frac{e^{3/4}GM}{2c_s^2}  \, ,
 \end{equation}
the numerical coefficient being almost exactly 1. The length-scale associated with the self-similar solution is now
 \begin{equation}
 R_S =   \frac{2 c}{A H} = \frac {8 c \, (\gamma -1)^{3/2}}{(3 \gamma -2)^{(3\gamma -2)/2(\gamma -1)}H}
 \end{equation}
and this is also shown in Fig.~\ref{scales}. Note that all three expression for $R_S$ shown there scale as $(\gamma -1)^{3/2}$ as $\gamma \rightarrow 1$ and the Babichev {\it et al.} expression actually coincides with the original Harada-Carr one in this limit. The ratio $HR_S/c$ is $0.5$ at $\gamma = 2$ and asymptotes to $8/(3\sqrt{3}) \approx 1.6$ as $\gamma \rightarrow \infty$, so the self-similar scale is always smaller than the separate-universe scale.

Clearly these equations do not apply for $\gamma <1$ since the value of $A$ given by Eq.~(\ref{bab2}) is not well defined then. However, Babichev {\it et al.} argue that one can take $A=4$ in this situation, corresponding to the stiff fluid value. In this case, Eq.~(\ref{bab1}) gives a sensible accretion rate, even though the accretion radius given by Eq.~(\ref{eq:radius}) is not well defined. We will use this result in the later discussion.
%Note that $K =\infty$ for $\gamma=0$, which gives zero critical mass. 
 
\subsection{Improved analysis allowing for cosmic expansion}
 
In all the above analyses, the existence of the self-similar black hole solution relates to the fact that the density scales as $\rho \propto t^{-2}$ in a flat Friedmann background. However, none of the refinements discussed above allows for the cosmological expansion itself,  which one might expect to oppose accretion, so one should be wary of the self-similar prediction. 

Carr and Hawking's refutation of the Zel'dovich-Novikov argument was based on the study of spherically symmetric self-similar solutions to Einstein's equations~\cite{ch1974}. Such solutions have the characteristic that all dimensionless quantities depend only on the  ``similarity'' variable $z \equiv r/t$, where $r$ is the comoving radial coordinate and $t$ is the cosmological time~\cite{ct1971}. The metric takes the form
\begin{equation}
ds^2 = -e^{2\Phi(z)} dt^2 + e^{2\Psi(z)} dr^2 + r^2S(z)^2 d \Omega^2,
\label{metric}
\end{equation}
where $d\Omega^2 \equiv d\theta^2 + \sin^2 \theta d\varphi^2$ and we now put $c=1$ throughout this paper unless specified otherwise. A key role is played by the ``velocity'' function 
\begin{equation}
V(z) = |z| e^{\Psi - \Phi} ,
\end{equation}
which represents the fluid velocity relative to the similarity surfaces of constant $z$ (which are expanding or collapsing spheres). Providing a certain regularity condition
 %(discussed later) 
is satisfied~\cite{hmc2006}, a value of $z$ where $V=1$ corresponds to a  null surface and is termed a ``similarity horizon''. 
The flat Friedmann solution is itself self-similar and has
\begin{equation}
V(z) \propto z^{(3\gamma -2)/(3\gamma)}\, .
%\quad S \propto z^{2/(3\gamma)} \, .
\end{equation}
The point where $V=1$ corresponds to the cosmological particle horizon for $\gamma > 2/3$. (It corresponds to a cosmological event horizon for $\gamma < 2/3$ but we assume $1<\gamma<2$ in this section.) If one had a black hole in a Friedmann background, there would be two values of $z$ for which $V=1$, the inner one being associated with the black hole event horizon and the outer one with the cosmological particle horizon. 

\begin{figure}[htbp]
\begin{center}
%\rotatebox{-90}{
\includegraphics[width=0.4\textwidth]{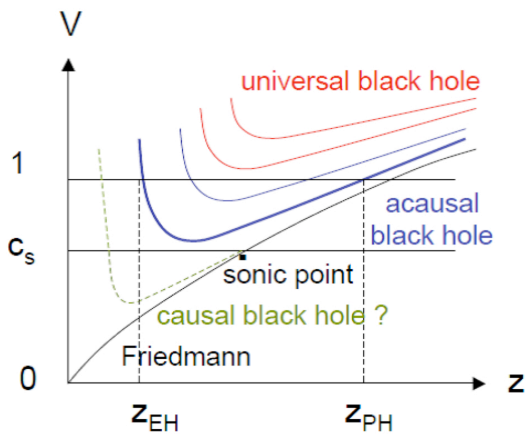}
%}
\caption{\label{velocity}
%$V$ is plotted against $z$ for $\gamma=1/3$.
Schematic figure of the function $V(z)$ for self-similar
asymptotically quasi-Friedmann black hole solutions, showing the two similarity horizons.
The solutions are described by a single parameter and the minimum of $V$ reaches the sonic value for solution. A black hole connected to an exact Friedmann background via a sound-wave would look like the dotted curve but such solutions cannot exist.}
\end{center}
\end{figure}

Values of $z$ for which $V=\sqrt{dp/d\rho}=\sqrt{\gamma -1}$ are also important because this may correspond to a sonic point (i.e. the pressure gradient may be discontinuous). For the radiation-dominated universe considered by Carr and Hawking, the sound speed is $V=1/\sqrt{3}$. Since the Friedmann solution itself has such a point, one might envisage a self-similar solution in which one attaches a black hole {\it interior} to an exact Friedmann {\it exterior} via a sound-wave. If such a solution existed, it would represent a black hole growing at the same rate as the universe and $V(z)$ would then have the form indicated by the broken curve in Fig.~\ref{velocity}. However, Carr and Hawking showed that  such a solution is not possible~\cite{ch1974}.  This implies that a black hole formed by local processes can never grow as fast as the universe, even if it starts off with a size comparable to the particle horizon. Thus the Zel'dovich-Novikov prediction cannot be correct in this case. 

This conclusion was soon supported by detailed hydrodynamical calculations. If the formation and evolution of a PBH in a positive-pressure fluid with a local perturbation is simulated numerically, 
%without assuming self-similarity, 
it is found that the PBH soon becomes much smaller than the
cosmological horizon~\cite{bh1979,nnp1978,np1980}.  In this situation, one {\it would} expect the Zel'dovich-Novikov argument to apply, so one does not anticipate the black hole growing very much at all. This also applies for holes which are initially larger than the horizon~\cite{hc2005c}, even though their mass should quickly diverge according to the Bondi-type analysis. 
%This is illustrated in the lower frame of Fig.~\ref{accrete}. 

On the other hand, Carr and Hawking also constructed a one-parameter family of spherically symmetric self-similar PBH solutions for a radiation ($\gamma=4/3$) or dust ($\gamma=1$) fluid which they described as {\it asymptotically} Friedmann~\cite{ch1974}. In these solutions,  as $z$ decreases, the velocity $V$ decreases to a minimum between $1$ and the sound-speed $1/\sqrt{3}$. $V$ then rises through $1$ at the black hole event horizon to infinity at the black hole singularity (corresponding to a finite value of $z$ but zero physical distance). Two such solutions are shown in Fig.~\ref{velocity}; 
%[REPEAT OF INTRO They correspond to special initial conditions, in which the matter is effectively thrown into the black hole at every distance. They do not contain a sonic point since they
we describe these as ``acausal'' since they are supersonic everywhere and
the perturbation extends beyond the cosmological horizon. There are
also solutions in which the minimum of $V$ exceeds 1, two examples of these being included. This corresponds to a ``universal'' black hole~\cite{cg2003} 
%[SUPERHORIZON BLACK HOLE?] 
in which there is no black hole or cosmological particle horizon.
%, although there are always both black hole and cosmological apparent horizons~\cite{c1976}. 
In this case, the entire universe might be regarded as being inside the black hole, although there are still black hole and cosmological apparent horizons~\cite{c1976}. 
%but they {\bf need not be observationally excluded.} 
%might still be physically plausible if the perturbation originated in the inflationary scenario.
The conclusion that there are self-similar black hole solutions in a universe which is asymptotically Friedmann but no such solutions in an exactly Friedmann universe was soon generalized
%and Bicknell and Henriksen~\cite{bh1978b} 
to perfect fluids with equation of state $p=(\gamma-1)\rho$ with $1 \le \gamma < 2$~\cite{c1976,bh1978b}.
%, where $\mu$ is the energy density~\cite{c1976,bh1978b}. 
%Indeed, it was found that no other interior solution can be attached to the exact Friedmann model apart from Friedmann itself: as $z$ decreases, either the mass goes negative or there is another sonic point at which the pressure gradient diverges~\cite{bh1978b}. [REPEAT OF INTRO]

It is interesting to examine these asymptotically Friedmann solutions in more detail. The single parameter $B_0$ which describes them is related to the perturbation at large $z$ of the factor $S(z)$ which appears in metric~(\ref{metric}):
\begin{equation}
S = S_{\rm F}(z) e^{B(z)}, \quad B \approx  B_0 +B_1 z^{-2(3\gamma -2)/(3 \gamma)},
\label{beta}
\end{equation}
where $B_1$ is determined by $B_0$ and the value of $\gamma$.
This is also related to the density perturbation at large distances:
\begin{equation}
W \equiv 8\pi G \rho r^2 = W_{\rm F}(z) e^{A(z)}, \quad A \approx  A_1 z^{-2(3\gamma -2)/(3 \gamma)},
\label{density}
\end{equation}
where $A_1 = (2-3 \gamma) B_1$.
For a sufficiently {\t negative} value of $B_0$, corresponding to a sufficiently large overdensity, $V$ will reach a minimum as $z$ decreases. This minimum 
will decrease as $B_0$ increases but it will exceed $1$ for $B_0 < \beta_1 < 0$, where the value $\beta_1$ depends on $\gamma$. These solutions correspond to universal black holes.

On the other hand, the minimum will lie between $V=1$ and $\sqrt{\gamma -1}$ for $\beta_1 < B_0 < \beta_2 < 0$, corresponding to the black hole solutions discovered by Carr and Hawking.  Here $\beta_2 $ also depends on $\gamma$ and is the value of $B_0$ for which the minimum of $V$ reaches $\sqrt{\gamma -1}$~\cite{cy1990}. Both these types of solution are also illustrated in Fig.~\ref{velocity}. Note that the ratio of the black hole event horizon size to the particle horizon size decreases as $B_0$ increases and so must always exceed the value associated with the solution with  $V_{\rm min} = \sqrt{\gamma -1}$. It might seem plausible to identify this with the Jeans criterion for gravitational collapse, in which case the ratio of the black event horizon and particle horizon should also have a minimum of order $\sqrt{\gamma -1}$. 
%Note that one does not need  as much fine-tuning of the black hole initial mass as in the Zel'dovich-Novikov analysis since these solutions are not exactly Friedmann and have an extra parameter.   

As $B_0$ increases above  $\beta_2$, the solution changes its form. Instead of $V$ having a minimum, the solution  reaches a sonic point, where 
%the solution can be discontinuous
%. However, if there is a sonic point, the solutions are  more complicated because 
the equations do not determine the behaviour uniquely, so 
there can be a discontinuity~\cite{cy1990}. 
In fact, only a subset of solutions are ``regular''  at a sonic point in the sense that  the pressure gradient is finite and they can be extended beyond there. There are two possible values of the pressure gradient, one associated with an isolated solution and the other with a 1-parameter family of solutions~\cite{OP}. These regular transonic solutions do not contain black holes. Indeed, one can prove~\cite{ch1974} that $V$ cannot have a subsonic minimum and then rise through another sonic point to reach $V=1$ again. Instead, they pass smoothly to the origin and represent either overdense perturbations (for $B_0 < 0$) or underdense perturbations (for $B_0 > 0$) which grow at the same rate as the universe~\cite{cy1990}. One can show that there is a continuum of underdense solutions and discrete bands of overdense ones~\cite{cy1990,OP}. The solution with $B_0=0$  is exactly Friedmann. These perturbed Friedmann solutions are of great interest in their own right~\cite{fog} but they are not relevant to the present considerations.

Since there is a one-parameter family of these solutions,  
we do not have to fine-tune the 
mass of the black hole to get 
self-similar growth as precisely as implied by Eq.~(\ref{eq:pbhmass}). This is not surprising since the earlier analysis assumed that the accretion was driven by a local sound-wave rather than by an initial perturbation extending to infinity. However, the range of the parameter $B_0$ for which self-similar growth is possible is still narrow and this reflects the proximity of the Jean length and  separate universe conditions. 

\begin{figure}[htbp]
\begin{center}
%\rotatebox{-90}{
\includegraphics[width=0.4\textwidth]{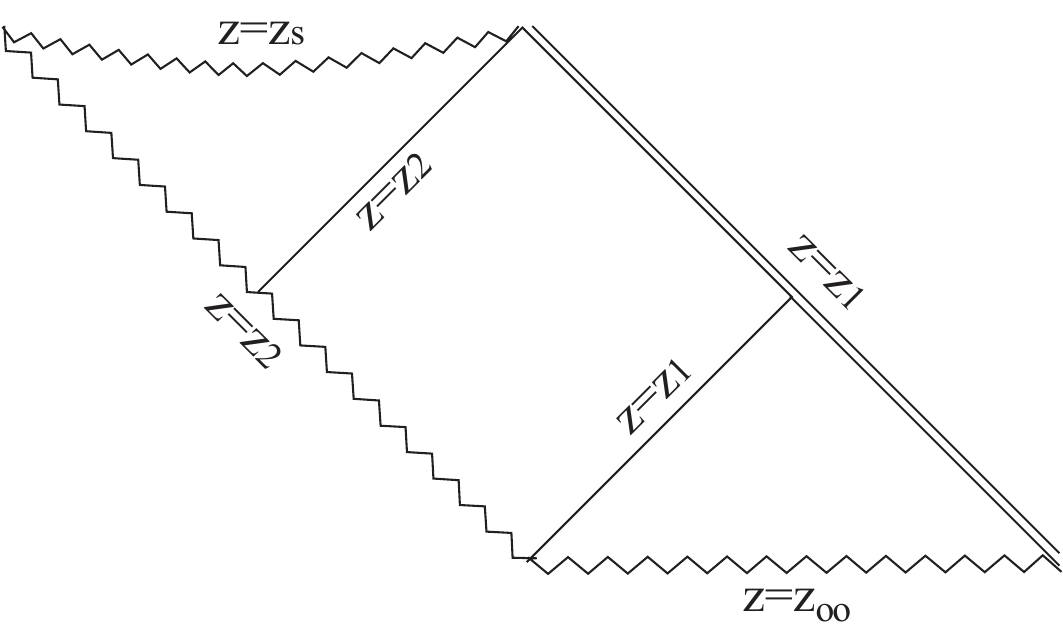}
\includegraphics[width=0.4\textwidth]{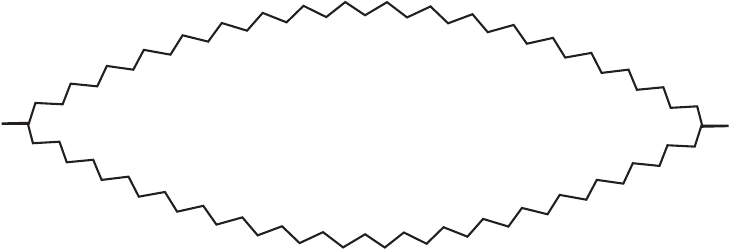}
%}
\caption{\label{penrose}
%$M$ is plotted against $t$ in the $\gamma=1/3$ case for the Newtonian (left) and relativistic (right) analyses.}
The Penrose diagrams for an acausal black hole (top) and a universal black hole full (bottom) in an asymptotically quasi-Friedmann background.}
\end{center}
\end{figure}

For comparison with the black hole solutions discussed later, Fig.~\ref{penrose} shows the Penrose diagrams for the acausal and universal black hole solutions. These correspond to Fig. 5 and Fig. 6 in Ref.~\cite{cg2003}, respectively.  In these (and subsequent) Penrose diagrams a zig-zag line corresponds to a singularity, while a double-line corresponds to infinity. In the acausal case, it is worth emphasizing that the conformal structure is very different from that for a non-self-similar black hole in a Friedmann background. In particular, the black hole singularity is connected to the big bang singularity and necessarily naked for a while. In the universal case, the conformal diagram has two spacelike singularities and there are no null infinities. This contrasts with the diagram for a recollapsing Fiedmann universe, which has two spacelike separated singularities. 

Later it was realized that these self-similar 
black hole solutions are not ``properly'' asymptotic to Friedmann after all~\cite{mkm2002}. The solid angle at infinity is  no longer $4\pi$ because the radial area -- measured by the quantity $S(z)$ -- is perturbed out to arbitrarily large distances. More precisely, Eqs.~(\ref{metric}) and (\ref{beta}) imply that the solid angle at infinity  is
$4\pi e^{2B_0}$, so there is a solid angle deficit in the black hole case ($B_0 <0$). These solutions may still be relevant to the real universe (since observations do not preclude such a deficit if it is not too large) 
%$B_0$ is close enough to $0$) 
but it would be more accurate to describe them as asymptotically ``quasi-Friedmann''. Note that one tends to the separate-closed-universe condition as $B_0 \rightarrow -\infty$ since the solid angle goes to zero. 
This is similar to the situation with the Barriola-Vilenkin monopole, which is an approximate general relativistic solution for the triplet Higgs scalar field in the region far from the center~\cite{bv1989}.
The spacetime is asymptotically flat in this monopole case but also contains a solid angle deficit.
The prospects of detecting such a deficit by gravitational lensing observations have been reviewed in Ref.~\cite{perlick2004}.

The complete classification of single-fluid shock-free positive-pressure spherically symmetric self-similar solutions by Carr and Coley~\cite{cc2000a} and Goliath et al.~\cite{gnu} has further clarified the situation. A key step in the Carr-Coley analysis is the derivation of all possible asymptotic behaviors at large and small distances from the origin. The behaviours at large spatial distances (usually corresponding to the limit $z \to \infty$) are of four kinds: (1) asymptotically quasi-Friedmann (1-parameter); 
%(2) asymptotically Kantowski-Sachs (but unphysical for $\gamma > 1$) (1-parameter); 
(2) asymptotically quasi-static (2-parameter). There are also two families of solutions which exist only when $\gamma > 6/5$: (3) asymptotically Minkowski at infinite $z$ (1-parameter); and (4) asymptotically Minkowski at finite $z$ but infinite physical distance (2-parameter). At small spatial distances, the solutions are also of four kinds: they contain either (a) a black hole or (b) a naked singularity at finite $z$; or they can be connected to the origin at $z=0$ via a sonic point, in which case they are either (c) static or (d) a perturbation of the Friedmann solution there. 

The complete family of $1<\gamma<2$ solutions is found by combining the four kinds of large-distance behaviour and four kinds of small-distance behaviour. The way in which one connects the large-distance and small-distance solutions depends crucially on whether or not there is a sonic point. 
If the solutions remain supersonic everywhere or subsonic everywhere, then the small-$z$ behaviour is uniquely determined by the large-$z$ behaviour. The asymptotically quasi-Friedmann black hole solutions are everywhere supersonic and of type (1a). However, if there is a sonic point, the solutions are  more complicated because
%the equations do not determine their behaviour uniquely there, so 
there can be a discontinuity. 
Indeed, we have seen that only a subset of solutions are regular  at a sonic point.
% in the sense that  the pressure gradient is finite and they can be extended beyond there. 
%These transonic solutions do not contain black holes; they pass smoothly to the origin and represent either underdense or overdense perturbations which grow at the same rate as the universe~\cite{cy1990}. 
Although we focus here on the black hole asymptotically quasi-Friedmann solutions, it should be stressed that these represent only a small subset of the total family of self-similar solutions. 

The study of critical phenomena in a Friedmann background~\cite{nj1999,jn1999,ss1999,hs2002,mmr2005} may also be relevant. If one considers the evolution of primordial density perturbations with a variety of forms, whose amplitude is described by some parameter $p$, one finds that a black hole forms if $p$ exceeds some critical value $p_*$ and that the black hole mass scales as
\begin{equation}
M \propto (p-p_*)^{\beta}
\end{equation}
for $p \approx p_*$. The exponent $\beta$ depends on the equation of state but is independent of the initial configuration of the collapsing fluid. The critical solution itself is self-similar and has been identified explicitly for different values of $\gamma$~\cite{ccgnu}. In terms of the Carr-Coley classification, it is either type (2d) (asymptotically quasi-static) for $1< \gamma < 1.28$  or type (3d) or (4d) (asymptotically Minkowski) for  $1.28< \gamma < 2$. (All the critical solutions have a regular centre and need to be asymptotically Friedmann there). Although the critical solution is only attained in the limit for which the black hole mass goes to zero,
% which might suggest that there can be no self-similar black hole solution with non-zero mass for this range of $\gamma$.  {\bf [TH: THIS IS WRONG. THERE ARE SUCH SOLUTIONS.]}. On the other hand, 
the critical solution is only an intermediate attractor, so there might still be a true self-similar attractor in which the mass is non-zero. In any case, we have seen that there are at least asymptotically quasi-Friedmann black hole solutions with non-zero mass for this range of $\gamma$.

%[This suggests that PBHs with non-zero mass cannot evolve self-similarly and the numerical studies support this conclusion. THIS ARGUMENT MAY NOT BE CONVINCING, ESPECIALLY SINCE THERE IS PROBABLY NO CRITICAL SOLUTION IN THE NEGATIVE-PRESSURE CASE, THERE BEING NO SOUND-WAVE AND ONLY ONE ATTRACTOR.] 
%{\bf [I DISAGREE THIS PART. THE TIME T FOR THE COLLAPSE TO A BLACK HOLE HAS THE OPPOSITE SIGN TO THAT FOR THE COSMOLOGICAL TIME IF WE CONSIDER  SELF-SIMILARITY. THE EXISTENCE OF CRITICAL PHENOMENA IN GRAVITATIONAL COLLAPSE HAS NOTHING TO DO THE VALIDTY OF SIMILARITY HYPOTHESIS FOR COSMOLOGICAL BLACK HOLES.]}
 
\section{Stiff fluid case}

For the limiting case in which $\gamma =2$, corresponding to a stiff fluid, the Zel'dovich-Novikov argument might conceivably apply because the sonic surface is a similarity horizon. Certainly this situation requires special
consideration. Lin {\it et al.} at first claimed that there {\it is}
a self-similar black hole solution in this case~\cite{lcf1976}. More precisely, they found a solution which has the exact Friedmann form outside the particle horizon (also a sonic point) but in which the velocity $V$ inside this point has a minimum and then rises to $1$ again, corresponding to a black hole event horizon.  Such a solution resembles the ones shown in Fig.~\ref{velocity}, except that it is subsonic between the event horizon and particle horizon. $V$ therefore has a subsonic minimum, which is precisely the situation Carr and Hawking excluded in the $1<\gamma <2$ case. The pressure gradient appears to diverge at the event horizon but Lin {\it et al.} argued that this divergence can be removed by introducing an Eddington-Finkelstein-type coordinate.

Later  Bicknell and Henriksen~\cite{bh1978a} showed that this solution is invalid because
the second point where $V$ reaches $1$ as $z$ decreases is not really a black hole event horizon. The reason for this is clarified in Ref.~\cite{hmc2006}. The metric induced on a constant-$z$ surface can be written as
\begin{equation}
ds^2 = -e^{2\Phi(z)}(1-V(z)^2)dt^2 + r^2S(z)^2 d\Omega^2,
\end{equation}
which is why $V=1$ is usually a null surface. However, although Lin {\it et al.} pointed out that the pressure gradient diverges where $V=1$, they failed to appreciate that the density goes to zero there and this implies that $e^{\Phi(z)} $ diverges. This means that the time component of the induced metric is non-zero, so the surface $V=1$ is timelike rather than null and cannot be a black hole event horizon. However, it is worth stressing that Lin {\it et al.} correctly deduced the form of the solution up to the second point where $V=1$.

Bicknell and Henriksen did manage to construct a numerical self-similar stiff solution containing a PBH
attached to an exact Friedmann exterior~\cite{bh1978a}. 
%However, in this solution  the stiff fluid turns into a null dust at a second similarity
%horizon (between the sound-wave and the black hole event horizon [{\bf
%Tomohiro and Hideki: This is not correct. The following is correct. 
However, this solution has the feature that the stiff fluid turns into null dust 
at the timelike surface where $V=1$, with the fluid becoming
lightlike there. This situation might appear to be contrived but  Reed and Henriksen~\cite{rh1980} were able to find a solution of this kind by generalizing a previous solution of Hacyan~\cite{hac1979} for a radiation-dominated model to arbitrary $\gamma$. In the Hacyan solution, the black hole is formed by inflowing radiation in a region described by the Vaidya metric.  Reed and Henriksen showed that this solution is necessarily self-similar, with the black hole mass growing asymptotically as $t$, but it is generally unphysical because it involves a decompression wave (which they claim violates the 2nd law of thermodynamics). However, they argued that it could be physical in the stiff case and that this might allow PBHs to form with masses up to $4 M_{\odot}$ if the equation of state is stiff up to the end of the hadron era. Although this seemed plausible at the time, the modern view does not favour a stiff hadron era, so this solution now also appears contrived. 

Recently, Harada {\it et al.}~\cite{hmc2006} have studied the stiff self-similar solutions with an exact Friedmann exterior in greater depth by treating the fluid equations as a 2-dimensional autonomous system and exploiting the equivalence (discussed below) between a stiff fluid and a massless scalar field. They find a family of extensions within the inner $V=1$ point which contain a massless scalar field.  These might appear more natural than the Bicknell-Henriksen or Reed-Henrksen solutions but they still enter a negative mass region, which is presumably unphysical. 

Harada {\it et al.} also prove that there is no asymptotically Friedmann or asymptotically quasi-Friedmann solution with a sonic point containing a black hole in the stiff case. This is analogous to the situation in the non-stiff case, where solutions with sound-waves can only represent  density perturbations at the origin rather than black holes. The possibility of self-similar solutions is even more restricted in the stiff case because there is no range for the minimum of $V$ between $1$ and the sonic value. However, as discussed below,
%recent work by Bhattacharya {\it et al.}~\cite{njb2010} suggests that 
one might still expect the existence of universal self-similar black hole solutions (with $V_{min} > 1$) in the stiff case, these being directly analogous to the supersonic solutions discussed in Section II.B. 
%[{\bf Kideki: I think we should be caferful not to give this comment since the stiff case is rather special and some miracle could  happen.} IS THE MIRACLE THE EXISTENCE OR NON-EXISTENCE OF SUCH A SOLUTION?] {\bf [TH: ACTUALLY, THERE ARE SUCH SOLUTIONS WITH STIFF FLUID AND THEREFORE MASSLESS SCALAR FIELD. I FOUND THE CONCRETE MODELS IN NAKAO'S TALK WORKSHOP TALK THIS MONTH BASED ON THE ONGOING WORK BY HIMSELF, JOSHI AND BHATTACHARYA.]}

\section{Scalar field case}

Scalar fields are a key ingredient  in modern cosmology.
They play an important role during 
inflation and certain phase transitions; they feature in the preheating scenarios; and they are pervasive in all sorts of alternative theories (eg. string theory and scalar-tensor theory) which are likely to be relevant in the high curvature phase of the early universe. 
In such scenarios, 
scalar fields necessarily dominate 
the energy density of the universe at some stage of its evolution. 
%This could have profound implications for the formation and evolution of PBHs.
One important physical feature in this case is that the scalar field can contain sound-waves, with the sound-speed ($\sqrt{ \delta p/ \delta \rho} \neq \sqrt{p/\rho}$) being the speed of light in the short-wavelength limit~\cite{hc2005b}. Equation~(\ref{eq:radius}) then suggests that the accretion radius is $\sqrt{2}$ times the Schwarzschild radius. As discussed later, this is different from the case of a perfect fluid with negative pressure but constant $\gamma$, in which there are no sound-waves. 

\subsection{Massless scalar field}

If there is no scalar potential, the scalar field is massless and one might expect the same results as in the stiff fluid analysis, since a scalar field is equivalent to a stiff fluid provided the gradient is everywhere timelike~\cite{m1988}. This is also consistent with numerical simulations of PBH growth in a Friedmann universe with a massless scalar field~\cite{hc2005c}. 
In fact, Harada {\it et al.}~\cite{hmc2006} have constructed an analytic proof that there is no self-similar PBH solution in an exact or asymptotically Friedmann background if the universe is dominated by a massless scalar field. The proof exploits Brady's 
formulation of the self-similar equations
as an autonomous system~\cite{brady1995} and uses Bondi coordinates $(v,r,\theta,\varphi)$ and the variables
\begin{equation}
y(\xi) \equiv 1-\frac{2M}{r}, \quad z(\xi) \equiv g^{-1}\frac{r}{v} \,.
\end{equation}
Here $v$ is the ingoing radial null coordinate, $M$ is the Misner-Sharp mass, $\xi \equiv \ln(r/|v|)$ and $g \equiv g_{vv}/g_{vr}$.

\begin{figure}[htbp]
\begin{center}
%\rotatebox{-90}{
\includegraphics[width=0.5\textwidth]{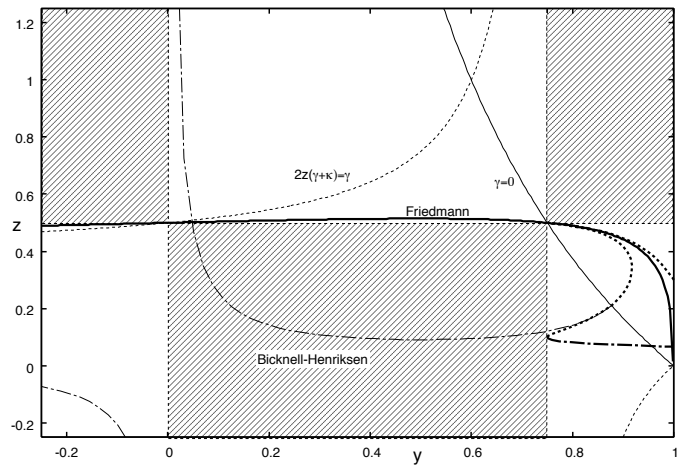}
%}
\caption{\label{yz_extended}
The massless scalar field solutions in the $(y,z)$ plane.
%, as described in Ref.~\cite{hmc2006}. 
%The shaded regions are prohibited for a real scalar field. 
The expanding flat Friedmann solution is plotted
as a thick solid line. 
%The gradient of the scalar field 
%can change from timelike to spacelike
%on $\gamma=0$ (thin solid line),
%$2z(\gamma+\kappa)=\gamma$ (thin
%dashed line) or  $y=0$, where the function $\gamma$ depends on  $y$ and $z$ [GIVE EXPLICTLY?].
%Note that the parts of the $2z(\gamma+\kappa)=\gamma$ and $\gamma=0$ curves in $z>1/2$ correspond to the null conditions for different solution branches  but are shown together for convenience. The flat Friedmann solution has a cosmological apparent horizon at $(y=0, z=1/2)$, a particle horizon at $(y=3/4, z=1/2)$ and ends at $(y=1, z=0)$.
Two numerical solutions within the particle horizon, whose significance is described in the text are 
shown by the thick dotted lines.
% and dashed-dotted line. 
% upper one 
%is a positive branch 
%goes directly to the negative mass region ($y>1$).
%The lower one reaches $y=3/4$
%, where it is converted from the positive branch to the negative branch 
%(shown by the thick dashed-dotted line) before going to the negative mass region.
The Bicknell-Henriksen null-dust black hole
solution
%, which connects to the lower numerical solution, 
is indicated by a thin dashed-dotted line.} 
%but this is never realized by a scalar field. 
%The black hole has an event horizon where the solution intersects $z=1/2$ and an apparent horizon at $y=0, z =\infty.$ } 
\end{center}
\end{figure}

The trajectories of the Harada {\it et al.} solutions in the $(y,z)$ plane are illustrated in Fig.~\ref{yz_extended}. The shaded regions are prohibited for a real scalar field. 
%The expanding flat Friedmann solution is plotted as a thick solid line. 
The gradient of the scalar field 
can change from timelike to spacelike
on $\gamma=0$ (thin solid line) or
$2z(\gamma+\kappa)=\gamma$ (thin
dashed line) or  $y=0$, where 
\begin{equation}
\gamma \equiv - \kappa \pm \sqrt {\frac{1+4 \pi \kappa^2 - y^{-1}}{4 \pi (1-2z)}} \, .
\end{equation}
%Note that the parts of the $2z(\gamma+\kappa)=\gamma$ and $\gamma=0$ curves in $z>1/2$ correspond to the null conditions for different solution branches  but are shown together for convenience. The flat Friedmann solution has a cosmological apparent horizon at $(y=0, z=1/2)$, a particle horizon at $(y=3/4, z=1/2)$ and ends at $(y=1, z=0)$.
Here $\kappa$ is a constant, which is taken to be $1/\sqrt{12\pi}$. Two numerical solutions within the particle horizon are 
shown:
% by the thick dotted lines, the dashed-dotted line
the one above the Friedmann line is the positive branch 
and goes directly to the negative mass region ($y>1$),
while the one below Friedmann reaches $y=3/4$
%, where it is converted from the positive branch to the negative branch 
%and the thick dashed-dotted line). 
 before going to the negative mass region.
The Bicknell-Henriksen null-dust black hole
solution connects to the lower numerical solution, 
%is indicated by a thin dashed-dotted line, 
but this is never realized by a scalar field. The black hole has an event horizon where the solution intersects $z=1/2$ and an apparent horizon at $y=0, z =\infty.$

Recent work by Bhattacharya {\it et al.}~\cite{njb2010} has found explicit self-similar asymptotically Friedmann black hole solutions with $V_{min} > 1$ in the massless scalar field case. These correspond to universal black hole solutions~\cite{cg2003} and are a special subset of the Roberts self-similar solutions~\cite{roberts}, the latter describing the inhomogeneous spherically symmetric  gravitational collapse of a massless scalar field which is minimally coupled to gravity (see also Ref.~\cite{ont1994}). Such solutions should therefore also exist in the stiff case. 

\subsection{Massive  scalar field}

The situation is more complicated -- but also more interesting -- if there is a scalar potential (i.e. if the scalar field is no longer massless). In particular, this applies in the quintessence scenario, in which the scalar field rolls down a flat potential~\cite{ratra}. 
%This has been invoked to explain the cosmic acceleration observed at the present epoch~\cite{supernova}.  
In this context, Bean and Magueijo~\cite{bm2002} have used a variant of the Zel'dovich-Novikov argument 
to claim that PBHs could grow up to $\sim 100M_{\odot}$ through the accretion of quintessence before nucleosynthesis and this could be large enough to 
provide the seeds for the supermassive black holes found in galactic nuclei.
A generalization of this analysis by Custodio and Horvath~\cite{ch2005}, though not involving self-similar growth, has also claimed there could be appreciable accretion of quintessence in some circumstances. 

We now examine these arguments in more detail. Factors of $c$ are included explicitly in this section for clarity and the dimensions of various quantities are justified in Appendix A. 
Bean and Magueijo~\cite{bm2002} consider a scalar field with a potential of the form
\begin{equation}
V(\phi)=V_{0} \,e^{-\sqrt{8\pi G} \, \lambda\phi/ c^2}
\label{eq:potential}
\end{equation}  
where $\lambda$ is a dimensionless constant. This is the only form of the potential compatible with a self-similar solution, so one can immediately exclude the possibility of a black hole growing as fast as the universe with any other form. Note that $\phi$ has units $c^2/ \sqrt{G}$ and $V$ has units $c^4/(GL^2)$ where $L$ is a lengthscale. 

By generalizing an analysis of Jacobson~\cite{j1999}, which attaches a Schwarzschild solution to a cosmological background in which the field has the asymptotic value $\phi_c(t)$, they claim that the energy flux through the event horizon is $\dot{\phi_c}^2/c$, where a dot denotes $d/dt$ and $t$ is cosmic time.
% [CONSTANT? {\bf NO!}], 
This leads to a black hole accretion rate~\cite{fk2003} 
\begin{equation}
\frac{dM}{dt} = 16\pi G^2M^2\dot{\phi_c}^2/c^7.
\label{eq:accretion2}
\end{equation}
They correctly note that only the kinetic energy of the scalar field is accreted. The scalar potential merely influences the evolution of the background field, the potential and the cosmic scale factor:
\begin{equation}
\phi_c = - \frac{2c^2}{\lambda \sqrt{8\pi G}} \ln\left(\frac{t_1}{t}\right), V = V_0 \left(\frac{t}{t_1}\right)^{-2}, a = \left(\frac{t}{t_1}\right)^{2/\lambda^2}
\label{background}
\end{equation}
where $t_{1}$ is defined by 
 \begin{equation}
t_1^2 \equiv  \frac{ c^2 (6-\lambda^2)}{4\pi GV_0 \lambda^4} \, .
  \end{equation}
%a constant of integration (the time at which $\phi_c$ is $0$). 
The expansion is accelerating providing  $\lambda < \sqrt{2}$. Since a scalar field has
 \begin{equation}
 \rho c^2 = \frac{1}{2c^2} \dot{\phi}^2 + V, \quad p = \frac{1}{2c^2} \dot{\phi}^2 - V \,,
  \end{equation}
%ndeed, since the above equations can be directly obtained using the Friedmann solution for a quintessence field, this verifies that the effective value of $A$ is 4. 
Eq.~(\ref{eq:accretion2}) can also be inferred from Eq.~(\ref{bab1}) of the Babichev {\it et al.} analysis if one assumes $A=4$. From Eq.~(\ref{eq:accretion2}) and the expression for $\phi_c$ given by Eq.~(\ref{background}), the accretion rate now becomes
\begin{equation}
\frac{dM}{dt} = \frac{KM^2}{t^2}, \quad K= \frac{8G}{c^3 \lambda^2} \, .
\end{equation}
%where we have assumed $A=4$ {\bf [hideki: 
%. so we should say ``it corresponds to $A=4$.''.]}. 
This equation can be integrated to give Eq.~(\ref{eq:pbhmass}) but with the revised expression for $K$.

As before, one has negligible growth for $M_0 \ll K^{-1}t_0$, 
self-similar growth for $M_0 \sim K^{-1}t_0$ and rapid divergence for $M_0 \gg K^{-1}t_0$. 
However, the self-similar mass 
is now of order the mass within the Hubble horizon (which is also the cosmological apparent horizon), since the particle horizon does not exist if the acceleration extends back to the infinite past.
%in the accelerating case ($\lambda^2<2$). 
% since the background universe has no particle horizon 
More precisely, using the scale factor given by Eq.~(\ref{background}), the Hubble mass can be shown to be 
\begin{equation}
M_{\rm H}=\frac{3c^3\lambda^2 t}{16 \pi G} = \frac{3t}{2K}\, ,
%\frac{2c^3\lambda^2 t_1^2}{G(2-\lambda^2)^3t} = \frac{2t_0}{K}\left( \frac{2}{2-\lambda^2}\right)^3 \left(\frac{t_1}{t} \right) \, ,
\end{equation}
%Please check my calculation in my note.]}
%[OK?] {\color{red}\bf [ROUGHLY OK]}.
which is just a factor $3/2$ larger than the mass corresponding to the self-similar solution.
%while the second factor is a number of order unity. 
%the difference second factor gives the correction to the self-similar mass $t_0/K$ at $t_0$
 %[BUT DIFFERENT FROM $K^{-1}t_0$?] 
The self-similar result is the one which Bean and Maguiejo exploit. 
However, their analysis is exactly equivalent to that of Zel'dovich and 
Novikov~\cite{zn1967}, which Carr and Hawking proved was inapplicable. It is therefore important  to know whether the non-existence of a black hole self-similar solution extends to this case.

Custodio and Horvath~\cite{ch2005} also consider black hole accretion of a quintessence field with a scalar potential but, instead of Eq.(\ref{eq:accretion2}), they use
 \begin{equation}
\frac{dM}{dt} = 27\pi G^2 M^2\dot{\phi_c}^2/c^7,
\end{equation}
%{\color{red}\bf [G FACTOR?]}
where the numerical factor differs 
%from that assumed in Eq.~(\ref{eq:accretion}) 
because of relativistic beaming. They disagree with Bean and Magueijo's choice of the function $\phi_c$ in Eq.~(\ref{background}) on the grounds that it neglects the local decrease in the background scalar field resulting from the accretion.  For reasons which are not altogether clear, they therefore focus on a model in which the quintessence flux into the black hole is constant. This requires that the potential have a specific form, which is different from (\ref{eq:potential}) and therefore incompatible with self-similarity. This leads to the mass evolution
 \begin{equation}
M=\frac{M_0}{1-\displaystyle{\frac{K' M_0}{t_0}}(t-t_0)} \, ,
\end{equation}
where $K'$ is a constant related to the (fixed) flux. This is
equivalent to the Zel'dovich-Novikov formula (\ref{eq:pbhmass}) providing one replaces $K$ by $K't$. In particular, the mass diverges at a time 
 \begin{equation}
t_{\infty} \equiv t_{0} \left( 1+ \frac{1}{K' M_0} \right)
\end{equation}
and they attribute this unphysical feature 
to the fact that the constant-flux assumption must fail. 
They therefore consider alternative models in 
which the flux decreases as a power of time. 
In general,  they find that the
increase in the black hole mass is small unless its initial value is
finely tuned.

Although Custodio and Horvath conclude that the Bean-Magueijo argument is
wrong, we would claim that  they have not identified the
more significant problem -- that it neglects the
background cosmological expansion.  
Indeed, since the cosmological expansion is neglected in both these analyses, they could both be flawed for the same reason as the original Zel'dovich-Novikov one.
%In any case, all these studies would appear to be inconsistent with our analytic proof that there is no self-similar PBH solution.

To examine the relativistic situation, we seek a spherically symmetric self-similar solution which represents a black hole in a quintessence-dominated cosmological background. In the quintessence scenario, for which the similarity assumption requires the potential to have the form given by Eq.~(\ref{eq:potential}), the analysis in Ref.~\cite{hmc2006} shows that there is no self-similar black hole solution in an exact or asymptotically Friedmann or asymptotically quasi-Friedmann background if the universe is decelerating and no such solution in an exact Friedmann background if it is accelerating. However,
this does not prove non-existence for the case in which the universe is accelerating and asymptotically Friedmann, which is what would be required to disprove the Bean-Magueijo proposal.  

Kyo {\it et al.}~\cite{khm2008} have shown that, 
with an accelerating scalar potential, there 
is a one-parameter family of self-similar solutions 
which are properly asymptotically Friedmann, in the sense that 
there is no solid angle deficit. 
In such solutions, the perturbation falls off very rapidly 
at spatial infinity. 
The question of whether such solutions can have a black hole 
event horizon remains open. However, as discussed in the next section, there is a self-similar black hole solution in a closely-related scenario. 
%[THIS MAY BE IRRELEVANT SINCE THERE ARE IMPORTANT PHYSICAL DIFFERENCES]

\section{Dark energy fluids}

%{\color{red}\bf [AS LONG AS I NOTICED,  THE TERMINOLOGY WAS NOT GENERALLY FIXED WHEN WE SUBMITTED THE "DARK ENERGY" PAPERS. BUT NOW I AM FEELING THAT IT HAS RECENTLY ALMOST COMPLETELY CONVERGED TO THE FOLLOWING: DARK ENRGY IS A MATTER FIELD WHICH CAUSES COSMIC ACCELERATED EXPANSION. THIS IS PARAMETRISED BY THE EQUATION OF STATE P=WRHO (W$<$-1/3). FOR -1/3$<$W$<$-1 IT IS CALLED QUINTESSENCE.FOR W=-1 IT IS CALLED COSMOLOGICAL CONSTANT. FOR W$<$-1 IT IS CALLED PHANTOM. QUINTESSENCE IS OFTEN REALIZED BY THE MODEL WHERE A SCALAR FIELD ROLLS DOWN ON THE VERY FLAT POTENTIAL. WE MAY CHOOSE TO USE THIS MODERN TERMINOLOGY.] }
The observed acceleration of the universe means that violation of the strong energy condition is required. The definition of the various energy conditions is given in Appendix B; for a perfect fluid with $p=(\gamma -1)\rho c^2$ and positive $\rho$, the violation of the strong one corresponds to $\gamma<2/3$ (including negative values).
% if the weak energy condition is satisfied. 
%[IS THIS TO EXCLUDE NON-NEGATIVE $\mu$?]
%{\bf [HM corrected and put an appendix for this part]}
We describe such a fluid as ``dark energy''. 
However, it should be stressed that this term is used in various ways by different authors and indeed is sometimes used interchangeably with the term ``quintessence'', so some clarification of terminology is required.  
In general, a scalar field corresponds to a stiff fluid ($\gamma =2$) when the potential is negligible and to a cosmological constant ($\gamma =0$) when the potential dominates, so it corresponds to a fluid with $0<\gamma<2$ at intermediate times
%. It can therefore be regarded as
and to  ``dark energy'' (in our sense of the term) for any period when $0<\gamma<2/3$. Quintessence corresponds to the special case in which the scalar field rolls down a very flat potential, so that $\gamma$ may change but is always in the latter range~\cite{ratra}.  
%The equation of state is usually written in the form $p = w \rho$, so $w \equiv  \gamma -1$ in our notation.  However, 
We have seen that there are sound-waves in the scalar field case but
%from the constant-$\gamma$ dark energy case is that there are sound-waves for a scalar field. The sound-speed $\sqrt{ \delta p/ \delta \rho}$ is different from $\sqrt{p/\rho}$ and equal to the speed of light in the short-wavelength limit~\cite{hc2005b}.  On the other hand, 
there are no sound-waves for a perfect fluid with $0<\gamma<1$ because $\sqrt{p/\rho} = \sqrt{\gamma -1} \, c$ is imaginary.

The question of whether a PBH can grow in a self-similar manner if the universe is dominated by dark energy can be addressed using the analysis of Babichev {\it et al.}~\cite{bab2004}.  
\if
For a general fluid, instead of Eq.~(\ref{eq:accretion}), one obtains~\cite{bab2004}
 \begin{equation}
\frac{dM}{dt} = \frac{4\pi G^2 AM^2}{c^3} \left( \rho_{\infty} + \frac{p(\rho _{\infty})}{c^2} \right),
\label{bab}
\end{equation}
where $A$ is a $\gamma$-dependent constant of order unity and the subscript $\infty$ indicates asymptotic values. For a positive-pressure fluid, this can be regarded as incorporating relativistic correction to Eq.~(\ref{eq:growth}). For a scalar field, one has
 \begin{equation}
 \rho = \frac{1}{2} \dot{\phi}^2 + V, \quad p = \frac{1}{2} \dot{\phi}^2 - V,
  \end{equation}
so Eq.~(\ref{bab}) reduces to Eq.~(\ref{eq:accretion2}). In the general case, gives Eq.~(\ref{eq:growth}) with $K$ is replaced by~\cite{bab2004}  
\begin{equation}
K = \frac{2A}{3G\gamma}.
\end{equation}
\fi
Although the parameter $A$ given by Eq.~(\ref{bab2}) is not well-defined for $0<\gamma <1$, Babychev {\it et al.} argue 
%(for reasons that are not altogether clear) [REALLY?] 
that it can be taken to be 4 throughout this range. In this case, Eqs.~(\ref{eq:growth}) and (\ref{bab3}) can still be applied, which suggests that self-similar growth is again possible. However, 
note that $K \rightarrow \infty$ as $\gamma \rightarrow 0$, so the self-similar black hole has zero mass in this limit.  
%[CF. CRITICAL PHENOMENA?]

As in the positive-pressure case, one should be wary of the self-similar claim since the above analysis neglects the cosmic expansion.  
The exact relativistic analysis has been carried out in our two recent studies of self-similar solutions with dark energy~\cite{hmc2007,mhc2007}. These use a combination of numerical and analytical methods to classify all spherically symmetric self-similar solutions with $0<\gamma<2/3$ which are asymptotically Friedmann at large distances. This extends the Carr-Coley classification of self-similar solutions with $1 < \gamma < 2$ to the $0 < \gamma < 2/3$ case. In fact, Ref.~\cite{cc2000b} already includes an asymptotic analysis of such solutions, although there are some errors in that work. 

There are several differences from the $1<\gamma \le 2$ situation. First, the fact that there are no sound-waves changes -- and indeed simplifies -- the analysis, since there can be no discontinuities and solutions are analytic everywhere. 
Second, the limiting values of $z$ for large and small spatial distances are reversed in the Friedmann solution: large spatial distances now correspond to $z \rightarrow 0$ and small ones correspond to  $z \rightarrow \infty$. Furthermore, we will find that some solutions necessarily span both positive and negative $z$ values.
There are also several differences as regards the solutions themselves. For example, one loses the exact static solution if $0<\gamma<2/3$ because one needs positive pressure to balance gravity. This also means that there are no asymptotically static ones, although there are still what are termed asymptotically quasi-static solutions (a point missed in Ref.~\cite{cc2000b}). On the other hand, one gains some solutions, since the Kantowski-Sachs and asymptotically Kantowski-Sachs 
%(i.e. with an angle deficit) 
solutions~\cite{ck1993} now become physical.
Although one loses the solutions for $6/5< \gamma < 2$ which are asymptotically Minkowski at {\it large} distances~\cite{cc2000a}, 
%these are replaced by 
one gains solutions which are asymptotic to a negative-mass Schwarzchild
singularity at {\it small} distances. 
%The crucial point is that there is still a 1-parameter family of solutions asymptotic to the flat Friedmann model at large distances (i.e. small $z$). However, there is still an important difference from the positive-pressure case. For $1<\gamma<2$, the solutions are only asymptotically {\it quasi}-Friedmann (because they exhibit an angle deficit at large distances) but they are genuinely asymptotically Friedmann in the $0<\gamma<2/3$ case.  

The $0<\gamma<2/3$ case is particularly important in the cosmological
context because there then exists a one-parameter family of self-similar solutions which are
properly asymptotic to the flat Friedmann model at large distances.
%, in the sense that there is no solid angle deficit at infinity. 
This has also been pointed out in Ref.~\cite{nusser2006}. As we have seen, this does not apply in the positive pressure case because the solutions are then only asymptotically quasi-Friedmann. However, these solutions are still acausal, in the sense that there is no sonic point, so they are not like the solutions which Carr and Hawking were originally seeking in the positive pressure context.

In order to investigate these solutions, we have used the asymptotic analysis of Ref.~\cite{hmc2007} and integrated the field equations numerically in Ref.~\cite{mhc2007}. We find that the solutions which are asymptotically Friedmann at large distances ($z \rightarrow 0$) are described by a single parameter $A_0$, which is analogous to the parameter $B_0$ in the  $1<\gamma<2$ case and may be interpreted as the density perturbation at spatial infinity. 
However, the expressions for $A$ and $B$ in Eqs~(\ref{beta}) and (\ref{density}) no longer apply and must be replaced by
\begin{equation}
A \approx  A_0 z^{(2-\gamma)/ \gamma}, \quad B \approx  -(A_0/6\gamma) z^{(2-\gamma)/ \gamma}
\label{alpha}
\end{equation}
%, which represents the density perturbation at large distances. 
in the $0<\gamma <2/3$ regime. The form of the solutions changes as one varies the parameter $A_0$ and one has three possible behaviours~\cite{mhc2007}.
The first class of solutions (with $A_0$ positive) is confined to the $z>0$ domain and contains a naked singularity; the Friedmann solution itself  has $A_0=0$. The second class (with $A_0$ negative but not too small) involves an extension into the $z<0$ region and contains a black hole; this is analogous to the
Carr-Hawking solution in the 
positive-pressure  case. The third class (with $A_0$ even more negative) also involves an extension into the $z<0$ region and corresponds to a wormhole solution; this has no analogy in the positive-pressure case.
%, although it is equivalent to the ``universal" black hole case.

\begin{figure}[htbp]
\begin{center}
\includegraphics[width=0.5\textwidth]{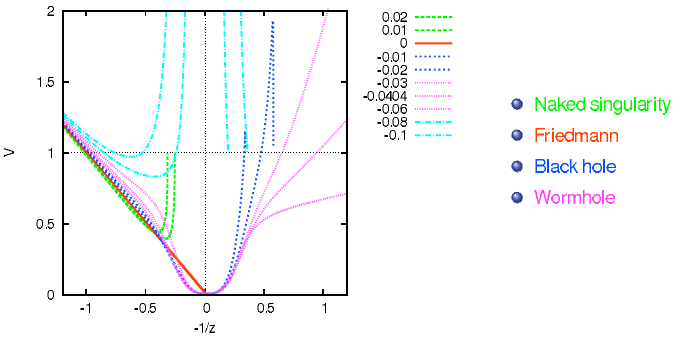}
\caption{\label{vel}
The form of $V$ in the negative pressure solutions with $\gamma = 1/3$, showing the transition from the black hole to wormhole to naked singularity solutions as one varies the parameter $A_0$ which describes their asymptotic behaviour. The ordinate is taken to be $-1/z$, which is large and negative in the asymptotic Friedmann region and small near the origin if the solution reaches there. }
\end{center}
\end{figure}

The form of $V$ in these solutions is shown as a function of $1/z$ in Fig.~\ref{vel} for the $\gamma =1/3$ case and we discuss them below in more detail. Note that $V=1$ now corresponds to a cosmological event horizon rather than a particle horizon. Although we did not investigate all values of $\gamma$ in the range $0$ to $2/3$, we would expect the results to be qualitatively similar.
We note that the system of equations was also numerically 
investigated in Ref.~\cite{nusser2006} but no black hole or wormhole solutions were reported there
because the analytic extension of the solutions to the $z<0$ domain was not considered.
%to the negative $z(\equiv r/t)$ is not considered in ref.~\cite{nusser2006}.
%As we will see in the present paper, this extension is crucial 

\subsection{Black hole solutions}

\begin{figure}[htbp]
\begin{center}
\includegraphics[width=0.4\textwidth]{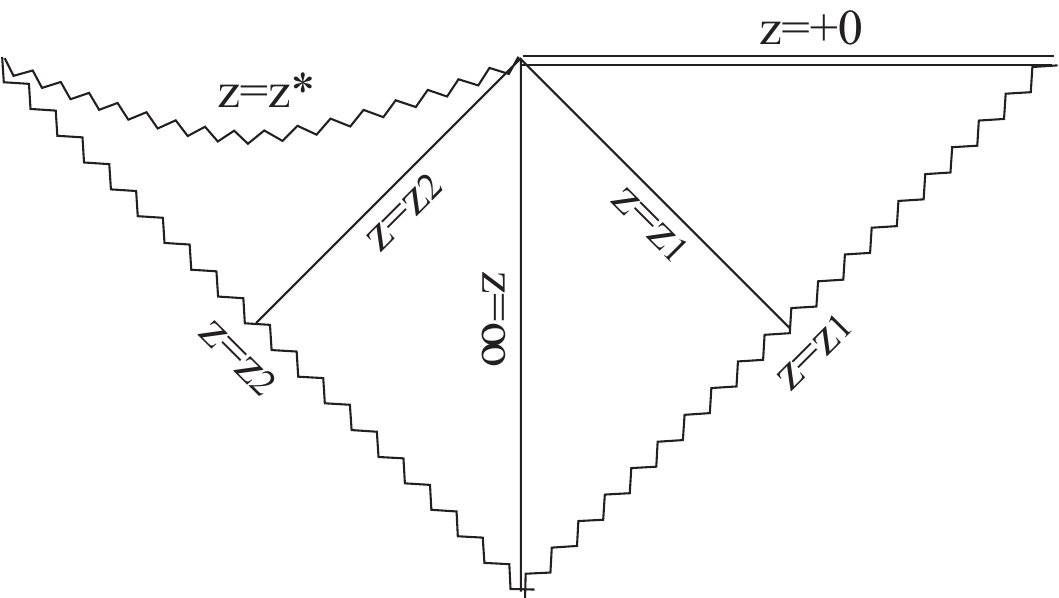}
\caption{\label{F-QKS-PMS}
The Penrose diagram for a solution with negative pressure which contains a black hole in an asymptotically Friedmann universe. }
\end{center}
\end{figure}

Solutions with $\alpha_{1} \le A_0<0$ for some critical $\gamma$-dependent value $\alpha_{1}$ (i.e. which are not too overdense) are asymptotically quasi-Kantowski-Sachs as $z \to \infty$. 
%(i.e. at the centre 
%{\bf [hideki: this is incorrect, it is neither physical center nor coordinate centor. Remove it.]}). 
The exact Kantowski-Sachs metric has the form
\begin{equation}
ds^2 = -\frac{(2-3\gamma)(2-\gamma)}{\gamma^2} dt^2 + t^{4(1-\gamma)/\gamma} dr^2 + t^2 
%(d\theta^2 + \sin^2 \theta d\phi^2)
d\Omega^2
%= d\theta^2 + sin^2 \theta d\phi^2
\end{equation}
with a suitable radial coordinate and this means that the physical distance tends to a finite limit (like a shell) rather than zero 
%{\bf at the centre (i.e. 
as $z \rightarrow + \infty$. 
%[hideki: ``at $z \rightarrow + \infty$'' is correct.]} . 
However, these solutions can be extended analytically into the region with negative $z$ and, as $z$ increases from $-\infty$, they then approach 
a positive-mass singularity at some negative value $z_{*}$. 
Since the Kantowski-Sachs solution has a curvature singularity at $t=0$, the analytic extension in this case must be interpreted as an extension from the positive $r$ to negative $r$ region. However,
the areal radius and mass are still positive in the extended region.
This solution describes a black hole in an asymptotically Friedmann background and is therefore analogous to the Carr-Hawking solution in the positive pressure case. In particular, there are two similarity horizons, 
corresponding to a cosmological event horizon at $z=z_1$ and a black hole event horizon at $z=z_2$, and a black hole singularity at $z=z_*$. However, 
%there are some differences:
the initial big bang singularity is null rather than spacelike in this case.
%, while the black hole singularity has both a null and spacelike portion {\bf [Hideki: I don't think so, since $z=z_{*}$ corresponds only to the spacelike one. ARE YOU SURE? HM: Yes, since $z=z_{*}$ corresponds to the BH singularity, please see fig.12 in Ref.~\cite{mhc2007} again.]}. 
The Penrose diagram for this solution is shown in Fig.~\ref{F-QKS-PMS} and is reproduced from Fig.12 of Ref.~\cite{mhc2007}.

For $\gamma=1/3$, we have found numerically that the ratio of the size
of the black hole event horizon to the Hubble length goes from $0$ and $0.70$ as $A_0$ decreases from $0$ to $\alpha_1$.
% [CF. CRITICAL SOLUTION HAS $0$ WHEREAS THERE IS LOWER LIMIT TO
%SIZE OF BLACK HOLE IN POSITIVE PRESSURE CASE. {\bf Hideki and Tomohiro: Yes. But the asymptotica srtucture is different from that case, so it is not so fare to compare these two cases.}]
This means that the size of a self-similar 
cosmological black hole has an upper limit, so only sufficiently small black holes 
can grow as fast as the universe, but may be 
arbitrarily small. Although a black hole may not grow as fast as the universe when it first forms (if it is as large as  the background Hubble length), the ratio of its radius
relative to the Hubble length will decrease as time proceeds. Eventually it will fall below  0.70, after which 
the black hole starts to grow self-similarly. 

Since there is a one-parameter family 
of these solutions,  
we do not have to fine-tune the 
mass of the black hole in the way implied by Eq.~(\ref{eq:pbhmass}) to get 
self-similar growth. 
% in the dark energy case. 
This also applied in the positive pressure case but the parameter range is now even more extended. Also the positive-pressure solutions may be less plausible since they are only asymptotically quasi-Friedmann.
%[FINE-TUNING IS MORE ASSOCIATED WITH FACT THAT THERE IS ONLY A NARROW RANGE OF SCALES WITH A LOWER LIMIT BECAUSE OF SEPARATE-UNIVERSE CONDITION, WHICH DOES NOT APPLY IF THERE IS AN UPPER LIMIT.]
%This can be explained that the pressure gradient of dark energy is negative, so that it gives an attractive force and pull the matter to the inside although gravitational force is not attractive but repulsive.

%The ratio of the black hole event horizon size to the cosmological event horizon
%size goes from $0$ to $0.36$, while
%the ratio of the black hole event horizon size to the 
%background Hubble horizon size goes from 0 to 0.70.

\subsection{Wormhole and white hole solutions}

\begin{figure}[htbp]
\begin{center}
\includegraphics[width=0.4\textwidth]{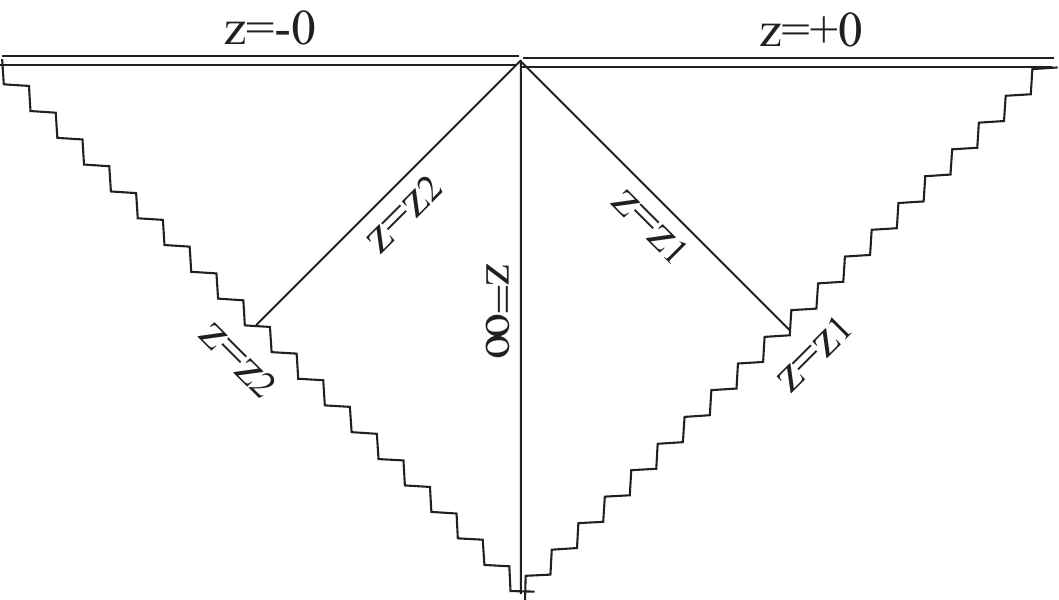}
\caption{\label{F-QKS-QF}
The Penrose diagram for solutions with negative pressure in which a wormhole connects an asymptotically Friedmann solution on the right to an asymptotically quasi-Friedmann universe on the left. }
\end{center}
\end{figure}

Solutions with $\alpha_{3} \le A_0<\alpha_{1}$ (where $\alpha_{3}$ is another $\gamma$-dependent critical value)
are also asymptotically quasi-Kantowski-Sachs as $z \to \infty$.
% {\bf (i.e. at the centre) [hideki:; remove]}. 
These solutions can again be analytically continued into the negative $z$ region but they now reach $z=0^{-}$ (corresponding to infinite physical distance) instead of a singularity. They are asymptotically Friedmann at $z=0^{+}$ but asymptotically quasi-Friedmann at $z=0^{-}$, except for a special value $\alpha_2$ between $\alpha_1$ and $\alpha_3$ which allows the solution to be asymptotically Friedmann at both ends. Since the physical  radius never becomes zero except at the big bang singularity, these solutions correspond to cosmological wormholes and there are two cosmological event horizons at $z_1$ and $z_2$.
%{\bf [hideki; should be ``Since the physical radius never becomes zero except for the big-bang singularity, these solutions correspond to cosmological wormholes'']}.
The Penrose diagram for these solutions is shown in Fig.~\ref{F-QKS-QF} and is reproduced from Fig.~15 of Ref.~\cite{mhc2007}. This only differs from Fig.~\ref{F-QKS-PMS} in that there is no singularity at the top left.

The transition from the black hole to wormhole solutions has a simple physical explanation. 
%As the parameter $A_0$ is increased, the ratio of the sizes of the cosmological and black hole apparent horizons decreases. 
As the parameter $A_0$ becomes more negative, the ratio of the size of the black hole apparent horizon to the size of the cosmological apparent horizon increases. 
The transition occurs when this ratio reaches unity and thereafter both horizons disappear, every region being trapped. This is in contrast to the $1 < \gamma < 2$ 
%{\bf [hideki: should be $1 \le \gamma < 2$ since the negative pressure case has not been confirmed.]}
case, where the two apparent horizons never merge and one tends to a separate closed universe as the black hole size increases~\cite{c1976}. 
% [EXPAND]. {\color{red}\bf [THIS EXAMPLE IS NOT FOR SELF-SIMILAR
%SOLUTIONS AND HENCE THE COMPARISON IS MEANINGLESS.]}
%{\bf [REALLY?] }
%The situations are contrasted in Fig.~\ref{wormhole}.
%\begin{figure}[htbp]
%\begin{center}
%\rotatebox{-90}{
%\includegraphics[width=0.5\textwidth]{separate_universe.eps}
%}
%\caption{\label{wormhole}
%This contrasts the form of the embedding diagram for a black hole, wormhole and separate universe .}
%\end{center}
%\end{figure}

\begin{figure}[htbp]
\begin{center}
\includegraphics[width=0.4\textwidth]{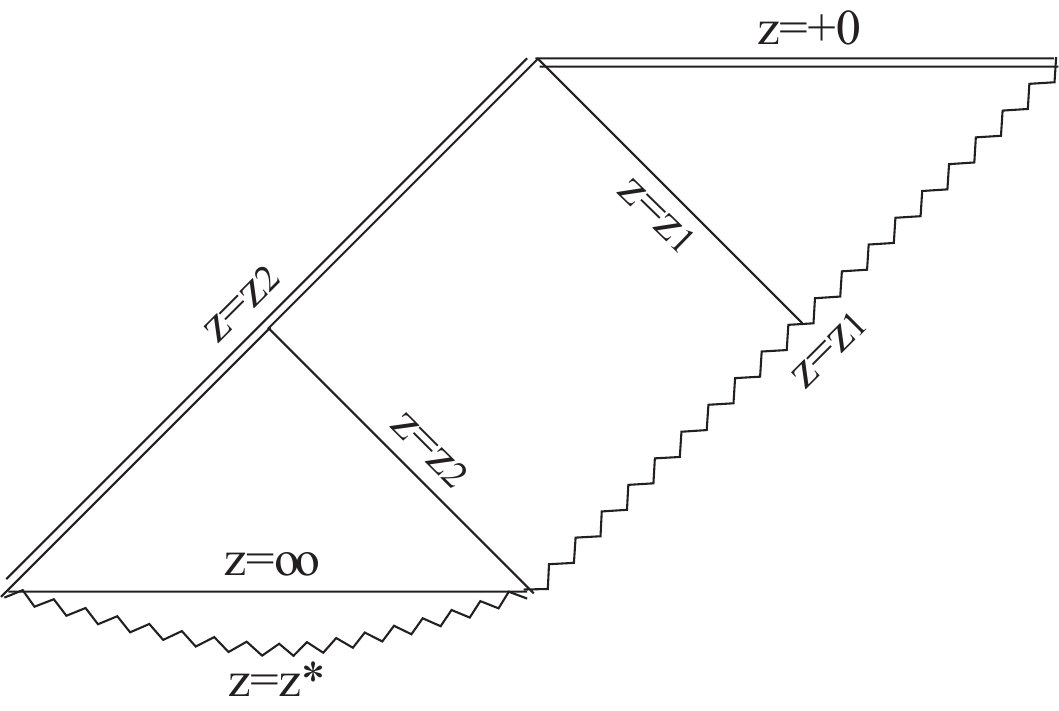}
\caption{\label{F-QS-PMS}
The Penrose diagram for a solution with negative pressure which describes the formation of a wormhole from a white hole in an asymptotically Friedmann spacetime. }
\end{center}
\end{figure}

The solutions with $A_0<\alpha_{3}$ are asymptotically quasi-static
rather than asymptotically quasi-Kantowski-Sachs as $z \to \infty$ but
they can still be extended into the negative $z$ region. These solutions
%also describe a wormhole but there is 
have a  central curvature singularity at some negative value $z_*$, where the mass is positive, and correspond to the formation of a wormhole from a white hole in a Friedmann background at $t=0$. 
% in the asymptotically quasi-static spacetime. 
%[SOLUTIONS END AT $V=1$?]
%{\bf [Hideki: I think it is better just to say ``They describe white holes in the Friedmann universe.'' since those white holes are eternal. THEN THIS SHOULD NOT BE UNDER WORMHOLE HEADING]} {\bf [TH: I AGREE.]}
%[DO WE NEED TO DISCUSS ASYMPTOTICALLY CONSTANT VELOCITY SOLUTIONS?
%{\bf NO! BECAUSE IT DOESN'T APPEAR IN NUMERICAL SOLUTIONS.}]
The Penrose diagram for this solution is shown in Fig.~\ref{F-QS-PMS} and is reproduced from Fig.15 of Ref.~\cite{mhc2007}. 

In order to understand these wormhole solutions, some clarification of what is meant by a wormhole is required. This is discussed in detail in another paper~\cite{mhc09} but we summarize the main points here.  A wormhole is an object in general relativity which connects two or more asymptotic regions.
The most famous example of a static wormhole is 
%first milestone came from Morris and Thorne. The so-called 
the Morris-Thorne solution~\cite{mt1988} and it 
is well known that this requires a violation of the null energy condition~\cite{visser,hv1997}. The definition of this is given in Appendix B but it corresponds to $\mu + p \ge 0$ for a perfect fluid.
%[GIVE EXPLICITLY? {\bf [hideki; Not necessary]}] 
%negative}. In the asymptotically flat case, this is also a consequence of the topological censorship~\cite{TC}. 
%[DEFINE IN APPENDIX] 
Dynamical wormholes are not so well understood but
their study was pioneered by Hochberg and Visser~\cite{hv1998} and Hayward~\cite{hayward1999}, who defined a wormhole throat in a dynamical spacetime
%two independent possible quasi-locally 
as some kind of trapping horizon~\cite{hayward1994}.
In their definitions, a wormhole throat is a two-dimensional surface of non-vanishing minimal area on a null hypersurface and the null energy condition must again be
violated there.
However, there is no past null infinity in the context of our cosmological wormhole solutions because there is an initial singularity. In fact, the Hochberg-Visser and Hayward definitions are inapplicable because the spacetime is trapped everywhere and there is no trapping horizon.
This demonstrates that these definitions miss
the important class of 
%dynamical wormhole spacetimes, namely 
 {\it cosmological} wormholes which are asymptotically Friedmann and start with a big bang.

In order to remedy this problem, we define a wormhole throat as a
two-dimensional surface of non-vanishing minimal area on a {\it spacelike} hypersurface~\cite{mhc09}. 
%The wormhole throats are therefore not defined on a null hypersurface but on a spacelike hypersurface. 
%so they are not Hochberg-Visser or Hayward dynamical wormholes.
%This motivates us to 
The one-parameter family of spherically symmetric self-similar wormhole solutions in an accelerating Friedmann background discussed above satisfy this definition. They are asymptotically Friedmann at one infinity and they have another infinity, which may also be asymptotically Friedmann 
for a special value of the parameter. 
%In this case, the wormhole throat connects two Friedmann universes.
Interestingly, the dominant energy condition 
%[GIVE EXPLICITLY? {\bf [hideki not necessary]}] 
is satisfied everywhere. 

In Ref.~\cite{mhc09} we
construct two analytic examples of
self-similar cosmological wormholes.
One corresponds to the numerical solution 
obtained above, but it contains a singular hypersurface which 
violates the null energy condition.
The other is a smooth model involving a combination of a perfect fluid and 
a ghost scalar field (i.e. a massless scalar field with a negative kinetic term) but the  total matter content still satisfies the dominant 
energy condition.

\section{Exotic cases}

In this section, we consider values of $\gamma$ which are outside the ``conventional'' range from $0$ to $2$ but might nevertheless arise in more exotic scenarios. These cases have not been studied in sufficient detail to come to definite conclusions. In particular, the spherically symmetric self-similar solutions have not been obtained explicitly.  Nevertheless, it is worthwhile summarizing what is known about these situations.

\subsection{Negative pressure fluids satisfying SEC}
A fluid with  $2/3 < \gamma< 1$ has negative pressure but it is not sufficiently negative to violate the strong energy condition. Self-similar solutions in this context have been studied in Ref.~\cite{cc2000b}, which shows that their possible large and small distance behaviours include asymptotically quasi-Friedmann and asymptotically Kasner  (corresponding to a black hole singularity), respectively. 
Indeed, the equations are formally similar to those in the $1 < \gamma< 2$ case, except for the absence of sound-waves and that should not be relevant if one is seekng an analogue of the $1 < \gamma< 2$ solutions which are supersonic everywhere. One would therefore {\it expect} the black hole solutions to exist, although this has not yet been demonstrated numerically.

\subsection{Phantom fluids}
A fluid with  $\gamma<0$ is described as a ``phantom''. More generally, this corresponds to any fluid with $p < - \rho$ and this possibility has been explored in the context of fluids with 
equation of state of the form $p = \alpha (\rho - \rho_0)$ in Ref.~\cite{bab2005}. Here we confine attention to the situation with $\alpha = \gamma -1$ and $\rho_0=0$ because of the similarly assumption.
%This also relates to the quintessence scenario:This means
%It therefore does not spend most of its time in this regime. 
%that there are important differences between the quintessence and dark energy scenarios.
%One could in principle extend the analysis to cover this case since 
There are still asymptotically Friedmann self-similar solutions in this case, so the previous analysis can in principle be extended to cover this. However, since the scale factor is $a(t)=|t/t_1|^{2/3\gamma}$, there is a ``big rip''
singularity~\cite{cald2002,cald2003} at $t=0$, spacelike infinity is at $z=+\infty$ and the domain of definition is $t<0$. Note that there is no future null infinity in these solutions, so there is no black hole event horizon. However, there is still a black hole apparent horizon and one may take this to define the mass of the black hole.
Equation~(\ref{bab1}) then has the consequence that the PBH
mass should decrease as a result of phantom accretion and it would need to decrease in proportion to the time left to the big rip in a self-similar solution.

Babichev {\it et al.}~\cite{bab2004} have analysed black hole accretion of 
phantom energy and find that the mass evolves according to
\begin{equation}
M=\frac{M_i}{1+ \displaystyle{\frac{M_i}{\dot{M_0} \tau} \frac{t'}{\tau - t'}}} \rightarrow \dot{M_0} |t| \, ,
%= \frac{M_i}{1+ \displaystyle{\frac{M_i }{\dot{M_0} |t| }} }
\label{phantom}
\end{equation}
where $M_i$ is the initial mass and $\dot{M_0} \equiv (3\gamma c^3 /2GA)$ with $A$ taken to be $4$. The first expression uses the Babichev {\it et al.} time coordinate with the time of the big rip corresponding to $t'=\tau$; the second expression applies in the limit $t' \rightarrow \tau$ and uses our time coordinate $t$. Note that Eq.~(\ref{phantom}) is very similar to  Eq.~(\ref{eq:pbhmass}) with $K$ being given by  Eq.~(\ref{bab3}).
%uses our time coordinate  with the big rip occurring at $t=0$. 
Although the mass evolution is clearly self-similar close to the big rip, the derivation of  Eq.~(\ref{phantom}) again fails to account for the background  expansion, since the black hole is not embedded in a cosmological background. Nevertheless, self-similar black hole solutions might in principle exist in this situation. The fact that black hole mass is {\it shrinking} rather than increasing is of obvious astrophysical interest. 
There have also been studies of black hole accretion in a ghost
condensate~\cite{frolov2004,mukohyama2005}. However, this term should not be confused with the word ``ghost'' which is sometimes used for the phantom fluid itself. A ghost condensate is more like dust ($\gamma =1$) than a phantom fluid. 
%. (A ghost condensate is like a phantom except that it is not a fluid [OK?].)  In these cases,

\subsection{Tachyonic fluids}
A fluid with $\gamma >2$ might be described as ``tachyonic'', in the sense that the sound-speed exceeds $c$. It formally contradicts the dominant energy condition but not the other ones. In fact, most of the equations in the $1 < \gamma< 2$ case be formally extended into this domain. In particular, all the curves in Fig.~\ref{scales} can be extended to arbitrarily large $\gamma$. Babichev {\it et al.}~\cite{bmv2008} have considered black hole accretion in this case but the problem does not yet seem to have been studied in the cosmological context.  

\section{Discussion and Summary} 

In this review, we have discussed the possible existence of self-similar
solutions containing a black hole or wormhole in an asymptotically Friedmann
background whose density is dominated by a perfect fluid with $p=(\gamma -1) \rho c^2$ 
%and $0<\gamma<2$ 
or a scalar field.
%we have also considered  the $\gamma<0$ and $\gamma>2$ cases. 
A simple Bondi-type analysis predicts an accretion rate of the form given by Eq.~(\ref{eq:growth}) in all cases, though with a different value of the constant $K$, which appears to permit self-similar growth. Ultimately, this is because the Friedmann equation implies that the density scales as $\rho \propto t^{-2}$ in a flat Friedmann background, which is precisely the condition for self-similarity. 

However, the simple analysis is suspect because it neglects the cosmic expansion, so this has motivated a more careful relativistic analysis which allows for the expansion. In the
positive-pressure case, the Bicknell-Henriksen solution (in which a
stiff fluid turns into null dust) seems to be the only known self-similar black
hole solution with a sound-wave which is {\it exactly} Friedmann at large distances and even this is rather contrived. There is also the Hacyan radiation-dominated solution (in which the region containing the black hole is described by a Vaidya solution) but this is also artificial because all the photons have to become radially directed within some point  and it also violates the 2nd law of thermodynamics. 

Apart from these examples,  it appears that there is no self-similar solution containing a
black hole in either an exact or asymptotically Friedmann background for any
value of $\gamma$ in the range $1 \le \gamma \le 2$. There are only asymptotically quasi-Friedmann solutions, including ``universal'' black holes without a black hole event horizon or cosmological particle horizon. However, there are self-similar asymptotically Friedmann solutions for $0<\gamma<2/3$, which
suggests that PBHs {\it can} grow as fast as the universe in the
presence of dark energy (at least for a limited period). This conclusion may also apply for a quintessence field, although this has not been rigorously proved. 

The difference between the positive and negative pressure solutions is important in two respects. First, while the negative-pressure ones  are physically well-motivated in the inflationary scenario, because one might expect the associated density perturbations to extend to ``infinity'',  the positive-pressure ones are theoretically unmotivated and may also be observationally excluded for some parameter because they have a solid angle deficit at large distances. Second, self-similar black holes only exist if their size as a fraction of the cosmological horizon is  not too small in the positive-pressure case but not too large in the negative-pressure case. This means that less fine-tuning is required in the latter case. It is interesting that there is no accretion in the limit $\gamma\to 0$, which is consistent with the 
Schwarzschild-de Sitter solution.

The existence of these self-similar black hole solutions suggests that  black holes may increase
their mass by a considerable factor during any dark-energy or quintessence dominated era, regardless of whether or not they are ``primordial''~\cite{bm2002} .  There are two contexts in which this effect may be important: (1) in the period immediately after any PBHs formed  at the end of inflation; (2) in the recent period when whatever dominates the density of the universe causes it to accelerate. Bean and Magueijo~\cite{bm2002} focussed on the first situation but the second one may also be interesting. For although the accretion factor may not be very large up to now, the black hole mass will continue to grow like cosmic time so as long as the dark energy 
dominates the cosmological density. In the simplest models, this applies indefinitely, so the black hole can grow arbitrarily large. 
We are investigating the astrophysical implications of this result -- and especially its implications for the Bean-Magueijo claim that PBHs can grow large enough to provide the supermassive black holes in galactic nuclei --  in a  separate paper.
%~\cite{chm2010}. 
\if
We have also considered the case of a phantom fluid with $\gamma<0$ and found that there may also be self-similar black hole solutions in this context, although we have been unable to come to a definite conclusion. The astrophysical implications of this result is of special interest since the black hole mass goes to zero. We have not yet considered the $\gamma>2$ case since it is not clear that this is physically well motivated. The other case which has been neglected is fluids with $2/3 < \gamma < 1$. However, we do not believe these are fundamentally different from $1< \gamma < 2$ as regards the accretion problem.
\fi

Finally, we note that our analysis restricts the situations in which the similarity
hypothesis~\cite{cc2000c} applies.
This hypothesis claims that there are certain circumstances in which spherically symmetric solutions evolve to self-similar form. The present work shows that there are at least some situations in which this does {\it not} happen. 
%Since this analysis shows that there is not always a self-similar solution, in the positive-pressure situation. This is because the hypothesis cannot hold if
(Even if there were a self-similar solution, one would still need to show that it was stable in order for it to be an attractor.) 
%he stability is important for the hypothesis because the existence does not mean that it
%is an attractor. Tomohiro: Add a reference like this.
%{\bf [I THINK THE ANALYSIS HAS NOTHING TO DO WITH THE GENERAL VALIDITY OF SIMILAITY HYPOTHESIS. WE JUST FIND IT IS NOT ALWAYS TRUE AND ACTUALLY THE HYPOTHESIS ALREADY INCLUDES THAT PHRASE IN ITS STATEMENT.]}
On the other hand, the similarity hypothesis {\it does} appear to hold in spherical 
gravitational collapse when the pressure is positive but very small 
$(0 < \gamma-1  \ll 1)$~\cite{hm2001,snajdr2006}. It may also hold in the negative-pressure situation. 

\acknowledgments
The authors would like to thank R.H. Henriksen, H. Motohashi and M. Nakashima for useful input. 
TH was supported by the Grant-in-Aid for Scientific
Research Fund of the Ministry of Education, Culture, Sports, Science
and Technology, Japan (Young Scientists (B) 18740144 and 21740190).
TH was also grateful to CECS
for its hospitality during his visit by the Fondecyt grant
7080214.
HM was supported by Fondecyt grant 1071125.
The Centro de Estudios Cient\'{\i}ficos (CECS) is funded by the Chilean
Government through the Millennium Science Initiative and the Centers of
Excellence Base Financing Program of Conicyt. CECS is also supported by a
group of private companies which at present includes Antofagasta Minerals,
Arauco, Empresas CMPC, Indura, Naviera Ultragas, and Telef\'{o}nica del Sur.
%TH was supported in part by JSPS Grant-in-Aid for Scientific Research No. 19340054 [OK?]
BJC thanks the Research Center for the Early Universe at the University of Tokyo for hospitality received during this work. This work has benefited from exchange visits supported by a JSPS and Royal Society  bilateral grant.

\appendix

%%%%%%%%%%%%%%%%%%%%%%%%%%%%%%%%%%%%%%%%%%%%%%%%%%%%%%%%%%%
\section{Physical dimensions}
%%%%%%%%%%%%%%%%%%%%%%%%%%%%%%%%%%%%%%%%%%%%%%%%%%%%%%%%%%%

We take the action to be
\begin{equation}
S=\int dtd^{3}x\sqrt{-g}\left[\frac{c^4}{16\pi G}{\ma R}-\left(\frac12 \phi_{,\mu}\phi^{,\mu}+V(\phi)\right)\right],
\label{action}
\end{equation}
where the coordinates in $4$-dimensional Lorentzian spacetime are $(x^0,x^i)=(ct,x^i)$ with $i=1,2,3$.
We emphasize that $dt \, d^{3}x$ must not be written as $d^4x$ if we include factors of $c$ explicitly.
The Einstein equations are
\begin{align}
G^{\mu}_{~~\nu}=&\frac{8\pi G}{c^4} T^{\mu}_{~~\nu}, \\
T_{\mu\nu}=&\left(\phi_{,\mu}\phi_{,\nu}-\frac{1}{2}g_{\mu\nu}\phi_{,\rho}\phi^{,\rho}\right)-g_{\mu\nu}V.
\end{align}
The 
%physical 
dimensions of the action and gravitational constant are
%has the same dimension of the angular momentum and hence
\begin{align}
[S]=ML^2T^{-1},\\
%\end{align}
%and the physical dimension of the gravitational constant is 
%\begin{align}
[G]=M^{-1}L^{3}T^{-2},
\end{align}
so we obtain
\begin{align}
[c^4/G]=&MLT^{-2},\\
[dt \, d^{3}x \times c^4/G]=&ML^{4}T^{-1},
\end{align}
and hence
\begin{align}
[T_{\mu\nu}]=ML^{-1}T^{-2}.
\end{align}
The relativistic factor
% following combination is seen 
 in the black hole metric is
\begin{align}
[GM/c^2]=L.
\end{align}

For a perfect fluid, the energy momentum tensor is 
\begin{align}
T_{\mu\nu}=(\mu+p)u^\mu u^\nu +pg_{\mu\nu},
\end{align}
where $u^\mu \equiv dx^\mu/d(c\tau)$ is the dimensionless 4-velocity of the fluid element, with $\tau$ being the affine time.
%, so $u^\mu$ is dimensionless.
%This is dimensionally correct since the physical dimensions for 
The mass density $\rho$, energy density $\mu$ and pressure $p$ have dimensions
\begin{align}
[\rho]=&ML^{-3},\\
[\mu]=&ML^{-1}T^{-2},\\
[p]=&ML^{-1}T^{-2},
\end{align}
so
%from which we have
\begin{align}
[G\mu/c^2]=[Gp/c^2]=[G\rho]=T^{-2}.
\end{align}
%The physical dimensions for the mass density $\rho$ is
%\begin{align}
%[\rho]=ML^{-3}.
%\end{align}
%Hence the linear equation of state can be written as $p=(\gamma-1)\mu=(\gamma-1)\rho c^2$.

For a scalar field, we have
\begin{align}
[\phi] =& [c^2/ \sqrt{G}] = M^{1/2}L^{1/2}T^{-1}, \\
%[G^{1/2}\phi]=& L^2T^{-2}, \\
%[G^{1/2}\phi/c^2]=&[0],\\
[V] =& [c^4/(GL^2)] = M L^{-1}T^{-2}.
\end{align}
Hence the exponential potential can be written in the dimensionally correct form 
\begin{align}
V(\phi) = V_0 e^{-\sqrt{8\pi G} \lambda\phi/c^2},
\end{align}
where $\lambda$ is a dimensionless constant.
The equation of motion for $\phi$ is 
\begin{align}
\dalm\phi=\frac{\partial V}{\partial \phi} \, .\label{kg}
\end{align}
In the cosmological context, a scalar field is equivalent to a perfect fluid with 
\begin{align}
\mu=&\frac12 \biggl(\frac{d\phi}{dx^0}\biggl)^2+V, \\
p=&\frac12 \biggl(\frac{d\phi}{dx^0}\biggl)^2-V.
\end{align}
%These are dimensionally correct.
For the flat Friedmann metric
\begin{align}
ds^2=-dt^2+(t/t_1)^{4/\lambda^2}dl_3^2,
\end{align}
where $dl_3$ is the flat line element, and 
\begin{align}
T^{t}_{~~t}=-\mu=-\frac{3c^2}{2\pi G\lambda^4 t^2}.
\end{align}
This is the energy density for a Friedmann universe filled by a scalar field with an exponential potential.
The last two equations also apply for a perfect fluid with $p=(\gamma-1)\mu$ if one replaces $\lambda$ with $\sqrt{3 \gamma}$.
% For the flat FRW metric
%\begin{align}
%ds^2=-dt^2+(t/t_1)^{4/(3\gamma)}dl_3^2
%\end{align}
 %so we obtain
%\begin{align}
%T^{t}_{~~t}=-\mu=-\frac{c^2}{6\pi G\gamma^2 t^2},
%\end{align}
%This is the energy density for the FRW universe filled by a perfect fluid with $p=(\gamma-1)\mu$.
%%%%%%%%%%%%%%%%%%%%%%%%%%%%%%%%%%%%%%%%%%%%%%%%%%%%%%%%%%%
\section{Energy conditions}
%%%%%%%%%%%%%%%%%%%%%%%%%%%%%%%%%%%%%%%%%%%%%%%%%%%%%%%%%%%

We summarize the energy conditions for a matter field with energy-momentum tensor
given in the diagonal form as
$T^\mu_{~~\nu}=\mbox{diag}(-\mu,p_{\rm r},p_{\rm t},p_{\rm t})$~\cite{he,carroll}.
The physical interpretations of $\mu$, $p_{\rm r}$ and $p_{\rm t}$ are the energy density, radial pressure and tangential pressure, respectively.

The inequality $T_{\mu\nu}W^\mu W^\nu\ge 0$ for any timelike vector $W^\mu$ is called the weak energy condition (WEC).
This guarantees that a timelike observer measures a non-negative energy density and
%.The weak energy condition (WEC)
 implies $\mu \ge 0$, $p_{\rm r}+\mu \ge 0$ and $p_{\rm t}+\mu \ge 0$.
The null energy condition (NEC) just replaces $W^\mu$ with a future-directed null vector in the above definition, which means that one can drop the condition $\mu \ge 0$.

The condition that $T^{\mu\nu}W_\nu$ is a future-directed non-spacelike vector for every future-directed non-spacelike vector $W^\mu$ 
%satisfying $T_{\mu\nu}W^\mu W^\nu\ge 0$ 
is called the dominant energy condition (DEC).
This means that the mass-energy can never flow faster than the speed of light and 
%the WEC, but the converse is not true.
%The DEC guarantees  in addition that the energy flux is a future-directed causal vector. 
%The dominant energy condition (DEC) 
implies $\mu \ge 0$, $-\mu \le p_{\rm r} \le \mu$ and $-\mu \le p_{\rm t} \le \mu$. 

The inequality $(T_{\mu\nu}-g_{\mu\nu}T^\rho_{~~\rho}/2)W^\mu W^\nu \ge 0$ for any timelike vector $W^\mu$ is called the strong energy condition (SEC).
If there is no cosmological constant, this assures the timelike convergence condition $R_{\mu\nu}W^\mu W^\nu \ge 0$ in general relativity, which 
means that the gravitational force is essentially attractive.
The SEC implies $p_{\rm r}+\mu \ge 0$, $p_{\rm t}+\mu \ge 0$ and $p_{\rm r}+2p_{\rm r}+\mu \ge 0$.

For $p_{\rm r}=p_{\rm t} \equiv p$, as assumed throughout this paper, the WEC implies $\mu \ge 0$ and $p+\mu \ge 0$, the DEC implies $\mu \ge 0$ and $-\mu \le p \le \mu$, the SEC implies $p+\mu \ge 0$ and $3p+\mu \ge 0$, and the NEC implies $\mu + p \ge 0$. The third and fourth conditions must be violated for inflation and the dynamical wormholes, respectively. Note that DEC implies WEC, WEC implies NEC, and SEC implies NEC, but SEC does not imply WEC.

%======================================%
%<<<<<<<<<<<<< REFERENCES >>>>>>>>>>>>>%
%======================================%

\end{document}